\documentclass[useAMS, usenatbib]{mnras}

\usepackage{verbatim}
\usepackage{appendix}
\usepackage{threeparttable}
\usepackage[all]{hypcap}
\usepackage{times}

\newcommand{\noopsort}[1]{}

\usepackage{newtxtext,newtxmath}

\usepackage[T1]{fontenc}
\usepackage{ae,aecompl}

\usepackage{graphicx}	
\usepackage{amssymb}	

\usepackage{epsfig}
\usepackage{subfigure}

\def \aap{A\&A }
\def \apjl{ApJ}

\def \apj{ApJ}

\def \jqsrt{J.~Quant.~Spectrosc.~Radiat.~Transfer}

\def \mnras{MNRAS}

\newcommand{\ts}{T_*}
\newcommand{\gs}{g_*}
\newcommand{\zs}{z_*}

\newcommand{\fE}{f_{\rm E,*}}

\newcommand{\fEc}{f_{\rm E,crit}}
\newcommand{\fEcr}{f_{\rm E,crit,rt}}
\newcommand{\fEcc}{f_{\rm E,crit,c}}
\newcommand{\fEs}{f_{\rm E,stab}}

\newcommand{\msun}{M_{\odot}}
\newcommand{\lsun}{L_{\odot}}

\newcommand{\fEavg}{\langle f_{\rm E}\rangle}

\def\alt{\raise0.3ex\hbox{$\;<$\kern-0.75em\raise-1.1ex\hbox{$\sim\;$}}}
\def\agt{\raise0.3ex\hbox{$\;>$\kern-0.75em\raise-1.1ex\hbox{$\sim\;$}}}

\newcommand{\bw}{\begin{widetext}}
\newcommand{\ew}{\end{widetext}}

\newcommand{\lsim}{\,\rlap{\raise 0.35ex\hbox{$<$}}{\lower 0.7ex\hbox{$\sim$}}\,}
\newcommand{\gsim}{\,\rlap{\raise 0.35ex\hbox{$>$}}{\lower 0.7ex\hbox{$\sim$}}\,}

\title[The Maximum Flux of Star-Forming Galaxies]{The Maximum Flux of Star-Forming Galaxies}

\author[R. M.~Crocker et al.]{Roland M.~Crocker,$^{1}$\thanks{E-mail: rcrocker@fastmail.fm (RMC)}
Mark R. Krumholz,$^{1}$
Todd A. Thompson$^{2}$
\newauthor
and Julie Clutterbuck$^{3}$
\\
$^{1}$Research School of Astronomy and Astrophysics, Australian National University, Canberra 2611, A.C.T., Australia\\
$^{2}$Department of Astronomy and Center for Cosmology \& Astro-Particle Physics, The Ohio State University, Columbus, Ohio 43210, U.S.A\\
$^{3}$School of Mathematical Sciences, Monash University, School of Mathematical Sciences
Monash University, Clayton 3800, Victoria, Australia
}

\date{Accepted XXX. Received YYY; in original form ZZZ}

\pubyear{2017}

\begin{document}
\label{firstpage}
\pagerange{\pageref{firstpage}--\pageref{lastpage}}
\maketitle

\begin{abstract}
The importance of radiation pressure feedback in galaxy formation has 
been extensively debated over the last decade. The regime of greatest 
uncertainty is in the most actively star-forming galaxies, where large 
dust columns can potentially produce a dust-reprocessed infrared radiation 
field with enough pressure to drive turbulence or eject material. Here we
derive the conditions under which a self-gravitating, mixed gas-star disc 
can remain hydrostatic despite trapped radiation pressure. 
Consistently taking
into account the self-gravity of the medium, the star- and dust-to-gas ratios,
and the effects of turbulent motions not driven by radiation,
we show that galaxies can achieve a maximum Eddington-limited star formation rate per unit area
$\dot{\Sigma}_{\rm *,crit} \sim 10^3 \msun$ pc$^{-2}$ Myr$^{-1}$, corresponding to a critical flux of $F_{\rm *,crit} \sim 10^{13} \lsun$ kpc$^{-2}$ similar to previous estimates; higher fluxes eject mass in bulk, halting
further star formation. 
Conversely, we show that in galaxies below this limit, our one-dimensional models imply simple vertical hydrostatic equilibrium and that radiation pressure is ineffective at driving turbulence or ejecting matter. 
Because the vast majority of star-forming galaxies  lie below the maximum limit for typical dust-to-gas ratios, we conclude that infrared radiation pressure is likely unimportant for all but the most extreme systems on galaxy-wide scales.
Thus, while radiation pressure does not explain the Kennicutt-Schmidt relation, it does impose an upper truncation on it. Our predicted truncation is in good agreement with the highest observed gas and star formation rate surface densities found both locally and at high redshift.
\vspace{.1in}
\end{abstract}





\noindent

\section{Introduction}
\label{sec:intro}

Young stars emit two thirds of their total energy and momentum budget at FUV and higher energies ($\gtrsim 8$ eV; \citealt{Krumholz2014}).
Such radiation interacts with dust grains with a large cross-section; even a galaxy  of moderate metallicity and gas column will absorb most photons in this energy range and, in so doing, reprocess them into the infrared. 
This process might be expected to affect the dynamics of the absorbing gas: the radiation field emitted by a zero-age stellar population carries a momentum flux per unit mass of stars formed of $\dot{V}_L \simeq 24$ km s$^{-1}$ Myr$^{-1}$ \citep{Murray2010,Krumholz2014b,Krumholz2014}.

While large enough to drive gas out of isolated proto-clusters experiencing intense star formation \citep{Krumholz2009a, Fall2010, Murray2010, Thompson2016}, this single-scattering radiation impulse is not large enough by itself to proffer a general explanation of the low star formation efficiency\footnote{Galaxies convert only $\sim 1$ \% of their cold gas to stars per free-fall time \citep[e.g.,][]{Zuckerman1974, Krumholz2007b, Krumholz2012b, Vutisalchavakul2016,Heyer2016, Leroy2017}.}
of galaxies on global scales:
the momentum budget  is simply too small \citep[][and references therein]{Andrews2011,Faucher-Giguere2013,Krumholz2014}.
However, because dust grains can accommodate a chain of multiple
scatterings or absorptions and re-emissions  for each photon emitted, 
significantly more momentum per unit time might, it seems, be extracted  from a light field by dust-bearing molecular gas than the $L/c$ obtained in the single-scattering limit.
Indeed, there have been suggestions  \citep{Thompson2005, Murray2010,Andrews2011, Hopkins2011} that, in the `strong trapping' limit, the rate of momentum deposition should approach $\sim \tau L/c$ which can considerably exceed the single-scattering value for large optical depths $\tau \gg 1$. 
Such large optical depths to reradiated infrared 
are encountered in the large gas column, dusty, star-bursting galaxies whose enormous, star-formation-driven 
radiative output emerges dominantly at long wavelengths \citep{Genzel2000,Calzetti2001}, with peaks at $\sim 100$ $\mu$m.
Such galaxies are detected locally as ultraluminous infrared galaxies (ULIRGs;
$L_\mathrm{IR} > 10^{12} \lsun$) like Arp 220 and as sub-mm galaxies at high ($z \gsim 2$) redshift.
Fits to these galaxies' observed spectral energy distributions imply dust columns of $\sim 0.01 - 0.5$ g cm$^{-2}$ \citep{Chakrabarti2008}, corresponding to optical depths of $\sim 10 - 100$ at 20 $\mu$m and $\sim 1 - 10$ at 100 $\mu$m.
  
In the strong trapping limit as may be applicable in such systems, 
amplification of the radiative momentum deposition is,
ultimately, limited only by energy conservation to $\lsim (c/v)L $ where $v$ is the characteristic speed of the outflowing gas \citep[e.g.,][]{Socrates2008}.
The claimed amplification has been controversial, however \citep[e.g.,][]{Krumholz2009a, Reissl2018}: in the optically thick limit, 
the gas distribution helps to shape the radiation (spatial) distribution and the coupling between gas and photons, mediated by dust, 
leads to various instabilities \citep{Blaes2003,Jacquet2011} whose impact must be assessed with numerical simulations.

Such considerations led \citet{Krumholz2012,Krumholz2013} to perform 
2-D direct radiation hydrodynamics simulations of gaseous discs subject to a constant, external gravitational field. They showed that the behaviour of gravitationally-confined, dusty columns of gas subjected to radiative fluxes is governed by two characteristic parameters: $\tau_*$, the dust optical depth, and the Eddington ratio, $\fE$, both computed for the opacity at the dust photosphere (see \autoref{eq_fE*} and 
\autoref{eq_tau*} below). Above a critical value of $\fE$ (to which we refer as $\fEcr$ below, and which depends on $\tau_*$), the gas cannot remain hydrostatic, and instead becomes subject to the radiation Rayleigh Taylor instability (RRT). This causes the gas to become turbulent and drives it into a density distribution that limits its ability to trap infrared photons. 

The simulations of \citet{Krumholz2012,Krumholz2013} used the Flux-Limited Diffusion (FLD) approximation to the radiation transfer problem, and later numerical studies employing more accurate radiative transfer approaches including the Variable Eddington Tensor  \citep[VET;][]{Davis2014,Zhang2017}, the implicit Monte Carlo radiation transfer \citep[IMC;][]{Tsang2015}, and the M1 closure \citep{Rosdahl2015, Bieri2017} schemes demonstrated significant differences in the behaviour of gas when $\fE > \fEcr$. In particular, the FLD simulations found that, above $\fEcr$, the gas becomes turbulent but is not ejected in a wind, while the VET and IMC studies find a continuous net acceleration of the gas that does launch a wind, though clumping does significantly reduce the acceleration of the wind compared to that which would be expected for a laminar matter distribution. In effect, the FLD simulations show that the ratio of the mass-averaged radiation force to gravitational force, $\fEavg$ approaches 1 from below while the VET and IMC simulations show that instead $\fEavg \to 1$ from above\footnote{The M1 simulations  find gas to be driven towards a velocity dispersion intermediate between the FLD and VET cases though also finding  that, as in FLD, the gas does not become unbound \citep{Rosdahl2015}}. Thus $\fE > \fEcr$ is a sufficient condition to guarantee that the atmosphere remains super-Eddington resulting in an outflow even though the gas is unstable, albeit one that accelerates much more slowly than suggested by early analytic estimates and subgrid models that did not consider the effects of RRT.

Conversely, for $\fE < \fEcr$ all numerical methods agree that the radiation drives no motion and injects no momentum. In this regime, dust-reprocessed radiation has no significant dynamical effects. The value of $\fEcr$ at which this transition occurs can be derived semi-analytically, without the need for any simulations at all. The existence of a critical value where trapped radiation sharply transitions from dynamically unimportant to capable of ejecting mass in bulk has important implications for its role in galaxy formation, which we explore in this paper. 
We extend the calculation of \citet{Thompson2005} and that of \citet{Krumholz2012} for the stability curve in a constant gravitational field to the more realistic case of a self-gravitating disc of mixed gas and stars. Based on our analysis we show that radiation pressure imposes an upper envelope on the range of gas and star formation surface densities, $( \Sigma_{\rm gas}, \dot{\Sigma}_* )$, that galaxies can explore. The vast majority of star forming galaxies, even luminous ones such as ULIRGs and sub-mm galaxies, lie far away from this envelope, and thus trapped radiation pressure cannot be responsible for regulating star formation in most galaxies or for determining the shape of the Kennicutt-Schmidt relation \citep{Kennicutt1998} on global scales. However, the upper limits of the observed galaxy distribution are intriguingly close to the calculated upper envelope, which strongly suggests that trapped radiation pressure does impose an upper truncation on the Kennicutt-Schmidt relation, and on galaxies' possible rates of star formation \citep{Thompson2005}.

The remainder of this paper is as follows. In \autoref{sec:setup} we introduce the basic equations that govern our model system, and in \autoref{sec:equilibria} we determine the conditions under which these equations admit stable and unstable equilibria. In \autoref{sec:implications} we consider the astrophysical implications of our findings, which we discuss further and summarise in \autoref{sec:discussion}. 
In a separate paper we will apply our results on indirect radiation pressure feedback
to star-forming sub-regions like individual giant molecular cloud complexes.

\section{Setup}
\label{sec:setup}

\subsection{Physical Configuration}

As a simple model of a galaxy disc we consider a planar distribution of stars and gas of infinite lateral extent with a vertical radiation flux $\mathbf{F}_* = F_* \, \hat{z}$  entering the domain of interest at $z=0$. By symmetry, we can just treat the half-plane from vertical height $z = 0$ to $z \to \infty$.
We assume that all radiation is injected at $z=0$ (i.e.\ there are no internal sources of radiation at $z>0$ except the thermal emission of the gas itself) and that there is local thermodynamic
equilibrium such that the dust, gas and radiation temperatures at any height $z$ are equal, $T_d(z) = T_g(z) =  T_r(z) \equiv T(z)$.
Note that the assumption of equal gas and dust temperatures is reasonable as long as the gas density is $\gtrsim 10^{4.5}$ cm$^{-3}$ \citep[e.g.,][]{Krumholz2014c} so that collisional coupling is efficient, which is the case for the starburst galaxies with which we are concerned.
 
In a steady state, the power entering the slab at $z=0$ must match the power escaping to $z \to \infty$.
\footnote{This is not true if enthalpy-bearing mass forms a wind that escapes to infinity; this is the so-called ``photon tiring" limit.
In \autoref{sec:tiring} we show that photon tiring is not a significant effect over the parameter space we consider though it may, given some generous assumptions, start to play a role 
for the very largest optical depths we investigate ($\tau_* \gsim 30$).} 
Moreover, as $z\rightarrow\infty$ the density must approach 0 for any physically reasonable configuration, so the gas must become optically thin at some sufficiently large $z$ (defining the photosphere).
This means that as $z\rightarrow\infty$ the flux and radiation energy density must approach the relationship 
$\mathbf{F}_\infty = c E_\infty \hat{z} = \mathbf{F}_*$. 
Following \cite{Krumholz2012} we  may thus define a reference temperature
\begin{equation}
\ts = \left(\frac{F_*}{ca}\right)^{1/4} \, ,
\end{equation}
where $a = 7.565 \times 10^{-15}$ erg cm$^{-3}$ K$^{-4}$ is the radiation density constant.
In steady state, $T(z)$ must approach $\ts$ as $z\rightarrow \infty$ and the free-streaming radiation energy density must satisfy $E_r = F_*/c \equiv E_*$.

We also define a reference acceleration at the top of the planar matter distribution
\begin{equation}
g_* = 4 \pi G \Sigma_{1/2} \, ,
\end{equation}
where $\Sigma_{1/2}$ is the column density in all matter (i.e., gas and stars) integrated outwards from the midplane 
(i.e., half the total column by symmetry; we use $\Sigma_{1/2}$ for the moment in order to ease comparison with the results of \citealt{Krumholz2012} 
whose setup was a  half-plane of gas in an external gravitational field).
In general we shall allow for both gaseous and stellar contributions to the total matter column as parameterized by the gas fraction
\begin{equation}
f_\mathrm{\rm gas} \equiv \frac{\Sigma_\mathrm{gas,1/2}}{\Sigma_{1/2}} = \frac{\Sigma_\mathrm{gas,1/2}}{\Sigma_\mathrm{gas,1/2} + \Sigma_\mathrm{stars,1/2}} \, .
\end{equation}
Note that only gas contributes to the integrated optical depth of a given column.

\subsection{Non-dimensionalisation}

With $\ts$ and $\gs$ in place we may now define a number of reference quantities, viz.:
i) the reference scale height of an isothermal gas distribution at $\ts$:
\begin{equation}
\zs =  \frac{k_B \ts}{\mu \gs} = \frac{k_B \ts}{\mu} \frac{1}{4 \pi G \Sigma_{1/2}} \, ,
\end{equation}
where $\mu$ is the mean molecular weight of the gas constituent particles;
ii) a reference gas density
\begin{equation}
\rho_* =  \frac{\Sigma_\mathrm{gas,1/2}}{\zs} = \frac{4 \pi G \mu \ f_\mathrm{\rm gas}  \Sigma_{1/2}^2}{k_B \ts} \, ;
\end{equation}
iii) a reference Eddington ratio (which is the Eddington ratio at infinity)
\begin{equation}
\fE =  \frac{\kappa_{R,*} F_*}{\gs c} = \frac{\tau_* F_*}{4 \pi G \ f_\mathrm{\rm gas} \Sigma_{1/2}^2 c} \, 
\label{eq_fE*}
\end{equation}
where we use iv) the reference optical depth (from the midplane to infinity but assuming the gas  temperature is fixed at the photospheric value):
\begin{equation}
\tau_* =  \kappa_{R,*} f_\mathrm{\rm gas} \Sigma_{1/2} \, ,
\label{eq_tau*}
\end{equation}
and v) the reference Rosseland mean opacity:
\begin{equation}
\kappa_{R,*} \equiv \kappa_R(\ts) \, .
\end{equation}
In physical units, 
\begin{equation}
\label{eq:opacity_scaling}
\kappa_R(\mathrm{10 \ K}) \sim 10^{-1.5} \ \mathrm{cm}^{2} \ \mathrm{g}^{-1} 
\end{equation}
for dust at Solar neighbourhood abundances; we assume that the dust abundance does not vary vertically within the gas column whose stability we are calculating, an assumption that is likely to be satisfied since turbulent motions will mix the dust vertically on the turbulent eddy turnover timescale.
Note that in general we can write
\begin{equation}
\kappa_R(T) = \kappa_{R,*} k_R(T/\ts) \, .
\label{eq_kappa}
\end{equation}
Given that the opacity of dusty material varies with temperature as roughly $\kappa \propto T^2$ (at temperatures $\la 150$ K; \citealt{semenov03a}), 
we will usually have below that $k_R$ in \autoref{eq_kappa} obeys
\begin{equation}
k_R(\Theta) =  \Theta^2 \, ,
\end{equation}
where here and in the following we use the  dimensionless temperature
$\Theta \equiv T/\ts$.
Below we also introduce a dimensionless, non-thermal ``temperature" parameter $\Theta_{\rm NT}$ to account for the fact that star-forming gas is extremely turbulent, with a non-thermal velocity dispersion that contributes an effective pressure that can be a significant, or even dominant, determinant of the overall dynamics.
Finally, we also define the dimensionless height
$\xi \equiv z/\zs$ and, using the column density given by
\begin{equation}
\Sigma(z) = \int^z_0 \rho\, dz  ,
\end{equation}
where $\rho$ is the gas density,
we can define
a dimensionless column density
\begin{equation}
s \equiv \frac{\Sigma}{\rho_* \zs} \, ,
\end{equation}
so that
\begin{equation}
\frac{d \Sigma}{d z} = \rho_* \frac{ds}{d \xi} \, .
\end{equation}

\subsection{Density and temperature profiles}

In order to be in mechanical equilibrium, we demand that there is momentum balance at all heights in the gas distribution.
Adopting the two-temperature flux-limited diffusion (FLD) approximation\footnote{Note that the limitations of FLD exposed by the VET and IMC treatments have to do with FLD's approximation of the radiation field direction
in a turbulent, porous flow. These limitations do not affect the hydrostatic case of interest here.} 
treated by \cite{Krumholz2007}, 
this condition of hydrostatic balance implies that 
\begin{equation}
\frac{d p_{\rm tot}}{d z} + \lambda \frac{d E}{d z} + \rho g = 0 
\label{eq_hydroBal}
\end{equation}
where $p_{\rm tot}$ is the total gas pressure (including a possible non-thermal pressure contribution from turbulence), $g$ is the height-dependent gravitational acceleration, and $\lambda$ is the dimensionless flux limiter which, following \citet{Krumholz2012}, we adopt from \citet{Levermore1981}  and  \citet{Levermore1984}.
Including both the gas self-gravity and an external gravitational field due to the stars, the expression for the $z$-dependent gravitational acceleration in our setup is
\begin{eqnarray}
g(z) & = &\int_0^z 4 \pi G \left[\rho(z') + \rho_s\right]\, dz' \nonumber \\
& \simeq & 4 \pi G \left[\Sigma(z) + \Sigma_{\rm stars,1/2} \right]  \, .  
\end{eqnarray}
Here the second near equality requires (as we shall henceforth assume) that 
the scale height, $z_s$ of the stellar volumetric density distribution  is much smaller than
the gas scale height 
so that we understand the gravity of the stars to be attributable to an infinitesimal mass sheet in the 
midplane.
This is a poor assumption for real galaxies, where the stellar scale height is always at least as large as the gas scale height. However, our goal here is to establish the position of the critical curve where hydrostatic equilibrium becomes impossible, which is defined by the divergence of the gas scale height. Thus in the locus we are interested in investigating, we are in fact in the limit where the stellar scale height is small in comparison to the gas scale height.

Note that in the FLD approximation (in one dimension) and given flux conservation, the radiation flux 
impinging from the mid-plane $F_*$ and the $z-$dependent radiation energy density $E = a T^4$ are related by
\begin{equation}
\lambda \frac{d E}{d z}  = -\frac{\kappa_R \rho F_*}{c} \, .
\label{eq_fluxCons}
\end{equation}
Using the invariance of flux with $z$ and directly adapting the results from \cite{Krumholz2012}, 
we can rewrite this equation as
\begin{equation}
\frac{a T_*^4}{\zs} \frac{d\left( \Theta^4 \right)}{d\xi} = -\frac{k_R \kappa_{R,*} \rho_*}{\lambda c} \frac{d s}{d \xi} \, .
\label{eq_fluxConsII}
\end{equation}

Finally, fully simplifying and using the scaling factors pre-defined above 
we have from \autoref{eq_hydroBal} and \autoref{eq_fluxConsII} a final pair of coupled, dimensionless,
ordinary differential equations (ODEs) that together specify the profiles of the dimensionless gas (surface) density, $s$, and dimensionless temperature, $\Theta$:
\begin{equation}
\frac{d}{d\xi}\left[\frac{d s  }{d \xi} (\Theta + \Theta_{\rm NT}) \right] = -\left[1 + f_\mathrm{\rm gas} (s -1) -  f_{E,*}  k_r \right]\frac{d s }{d \xi} \, ,
\label{eq_mtmBalI}
\end{equation}
and
\begin{equation}
\frac{d}{d\xi} \left(\frac{\lambda \Theta^3}{k_r} \frac{d \xi}{d s} \frac{d \Theta}{d\xi} \right) = 0 \, ,
\label{eq_tempGrad}
\end{equation}
where in
\autoref{eq_mtmBalI} 
we have accounted for turbulence via the introduction of the (constant) non-thermal temperature $\Theta_{\rm NT}$ (which, note, is absent from \autoref{eq_tempGrad}). We assume that $\Theta_{\rm NT}$ is independent of $\xi$, which amounts to assuming that the velocity dispersion that characterises the turbulence is independent of height. This assumption is consistent with observations, which do not show large vertical gradients in ISM velocity dispersions. It is also what is expected from the fact that turbulent motions generally have most of their power on the largest scales, which implies that most of the support is provided by motions on size scales comparable to the gas scale height, precluding the possibility of variations on smaller scales.
Physically, \autoref{eq_mtmBalI} 
above asserts that the gas pressure gradient balances the force of 
gravity (from gas and stars), diluted by radiation pressure, at every point; \autoref{eq_tempGrad} asserts that the temperature gradient is such as to conserve the radiation flux \citep{Krumholz2012}.

Adopting the results of \citet{Krumholz2012} the boundary conditions (BCs) for this ODE system are
\begin{eqnarray}
s(0) & = & 0 \label{eq_BC4} \\
\lim_{\xi \to \infty} s(\xi) &=& 1 \label{BC_1} \\ 
\lim_{\xi \to \infty} \Theta(\xi)  &=& 1  \label{BC_2}\\ 
\left.\frac{d\Theta}{d\xi} \right|_{\xi=0} & = & -\left.\frac{\tau_* k_R }{4 \lambda \Theta^3} \frac{ds}{d\xi} \right|_{\xi=0} \label{BC_3}  \, .
\end{eqnarray}
Here BC~\ref{BC_1} is equivalent to demanding that $\int_0^\infty \rho\,dz = \Sigma_{1/2}$ and BCs~\ref{BC_2} and \ref{BC_3} are equivalent to demanding that the flux be $F_*$ 
as $z \to \infty$ and
at $z=0$, respectively.
Note that the gas density gradient at the midplane $\xi = 0$, where the gravitational field due to the gas vanishes, obeys the equation:
\begin{equation}
\left.\frac{d^2 s}{d\xi^2}\right|_{\xi=0} = - \left[\frac{d s}{d\xi} \frac{1}{\Theta + \Theta_{\rm NT}}   \left(1- f_\mathrm{\rm gas} - f_{E,*} k_r  + \frac{d\Theta}{d\xi} \right)\right]_{\xi=0} \, .
\label{eqn_MPdensGrad}
\end{equation}

\begin{figure}
\includegraphics[width = \columnwidth]{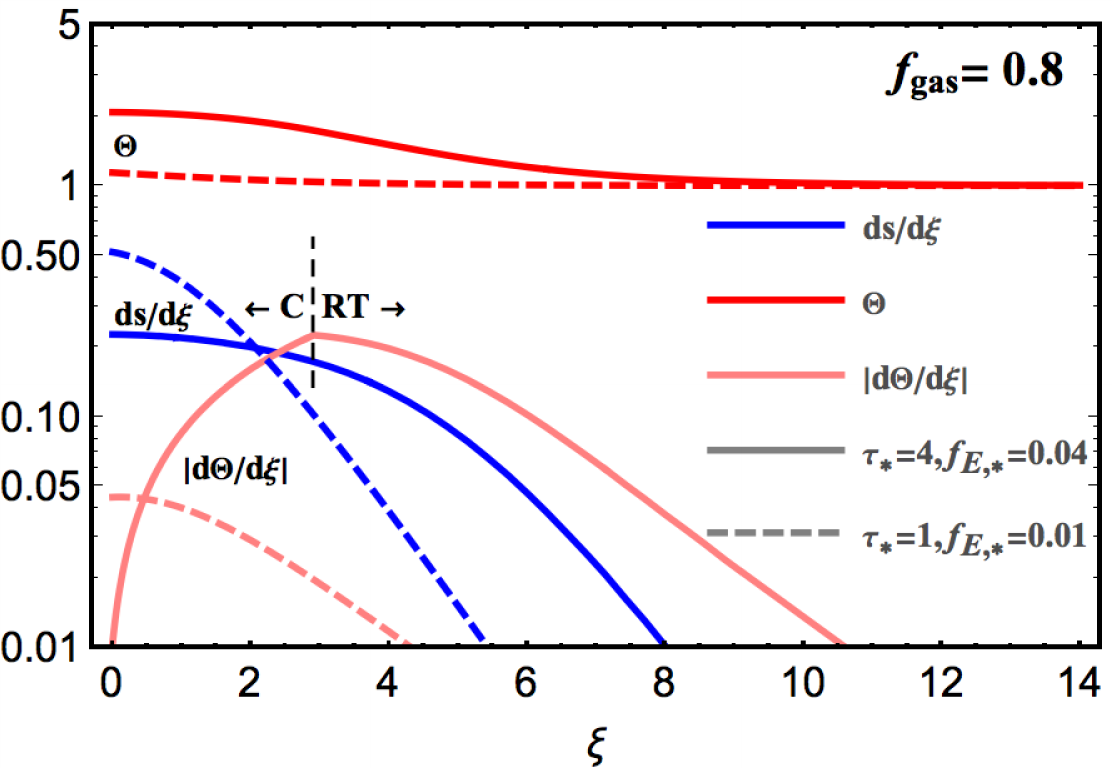}
\caption{Equilibrium profiles for the 
dimensionless volumetric density $ds/d\xi$ (blue),  dimensionless temperature $\Theta$ (red), and absolute value of the dimensionless temperature gradient
$d\Theta/d\xi$ (pink)
in a self-gravitating, gaseous disc with $f_{\rm gas} = 0.8$
computed for an opacity law $k_R = \Theta^2$ and with $\fE = 0.04$, $\tau_* = 3$ (solid) and $\fE = 0.01$, $\tau_* = 1$ (dashed).
For the latter (dashed line) case, the temperature profile is governed by radiative transfer exclusively; 
for the former (solid line), the temperature profile is governed by convection (i.e, \autoref{eq_tempGradCSII}) for $0 \leq \xi < 2.9$ (`{\bf C}') and by radiative transfer (i.e., \autoref{eq_tempGrad}) for $\xi \geq 2.9$ 
(note the kink in the solid pink curve at $\xi = 2.9$). 
The Eddington ratios at $\xi = 0$ for these solutions are 0.17 and 0.013, respectively.%
}
\label{fig_plotProfilesForVADERPaper1}
\end{figure}

\autoref{fig_plotProfilesForVADERPaper1} shows some example dimensionless temperature and volumetric density ($ds/d\xi$) profiles obtained by solving our ODE system numerically with $\Theta_{\rm NT} = 0$. 
One example shows a case that is entirely convectively stable, while the other shows a case where convection occurs (see \autoref{sec:equilibria}).
 
\section{Equilibria}
\label{sec:equilibria}

In this section we first consider the case of no turbulence, $\Theta_{\rm NT} = 0$, before exploring the effects of non-zero $\Theta_{\rm NT}$ in \autoref{sec:ThetaNT}.

\subsection{Existence of equilibria}

\begin{figure}
\centering
\includegraphics[width = \columnwidth]{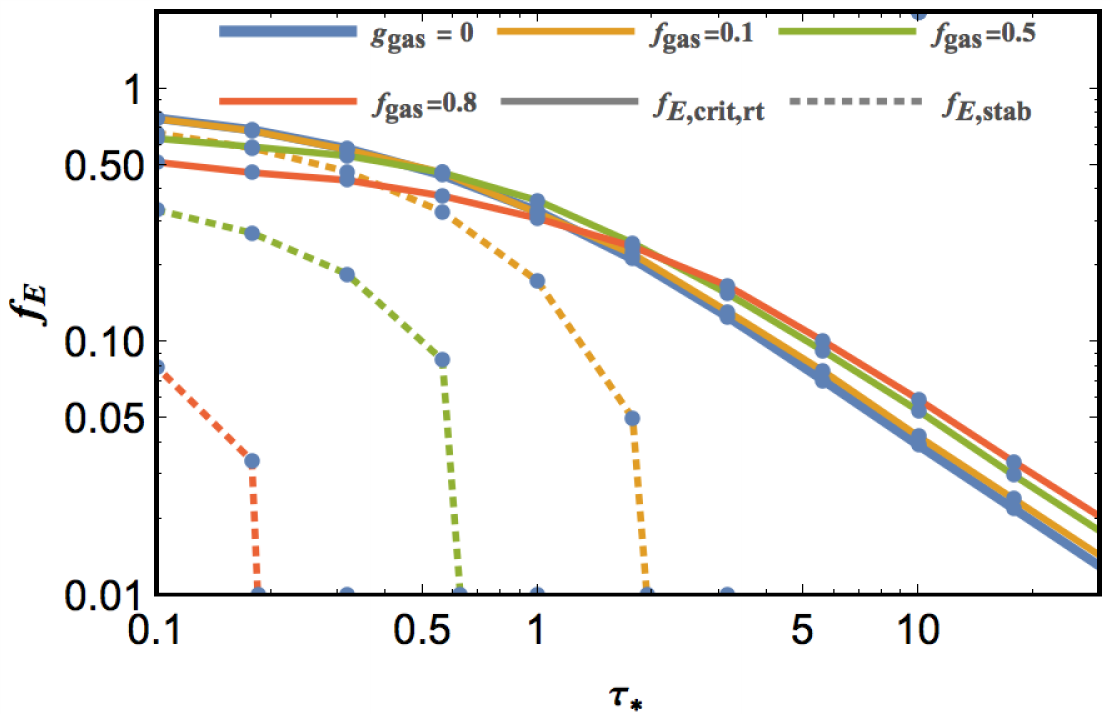}
\caption{
Curves for $\fEcr$ (solid) and for $\fEs$ (dotted) at different gas fractions, as labelled, with $k_R =  \Theta^2$ and $\Theta_{\rm NT} = 0$. 
The case labelled $g_\mathrm{gas} =\mbox{0}$ corresponds to vanishing gas self gravity (while maintaining finite optical depth) 
as investigated by \citet{Krumholz2012};
this case is obtained by setting $f_{\rm gas} = 0$ in \autoref{eq_mtmBalI} (and \autoref{eqn_MPdensGrad}).
The solid curves for $\fEcr$ show the
largest $\fE$ that permits equilibrium for a column whose temperature profile is governed by radiative transfer without regard to convective stability.
The dotted curves for $\fEs$ show the
maximum values of Eddington ratio at given $\tau_*$ such that convectively stable equilibria are possible.
} 
\label{fig_plotStabilityRegion}
\end{figure}

An important feature of the system of equations we have written down is that the existence of an equilibrium solution for an arbitrary combination of $\fE$, $\tau_*$, and $k_R$ is not guaranteed; rather, for any specified $\tau_*$, and $k_R$ 
there will exist a critical  Eddington ratio, called by us $\fEcr$ (with r.t. = radiative transfer)\footnote{Labelled $\fEc$ by \citet{Krumholz2012}.} 
above which radiation pressure is too strong for a hydrostatic atmosphere to form. In this case the gas becomes turbulent as a result of radiation Rayleigh Taylor instability, and may become unbound entirely.
A necessary but  insufficient condition to guarantee mechanical equilibrium can be obtained by consideration of the $\xi \to \infty$ limit of \autoref{eq_mtmBalI}: for finite $\Theta \to 1$ at $\infty$ from BC.~\ref{BC_2}, we have that $k_R \to 1$ at $\infty$; a finite gas column (BC.~\ref{BC_1}) then requires that   $1 -  f_{E,*}$ be positive.
However, even if $\fE < 1$, there may still be no solution that obeys both BCs \ref{BC_1} and \ref{BC_2}.
Operationally, we follow \citet{Krumholz2012} by using a shooting method 
to numerically determine $\fEcr$, the maximum  value of $\fE$ for which an equilibrium solution exists for a given optical depth $\tau_*$ and for $k_R =  \Theta^2$.
We show $\fEcr$ as a function of $\tau_*$ at sample values of $f_{\rm gas}$ in \autoref{fig_plotStabilityRegion} (solid lines). 
For comparison we also show the value of $\fEcr$ obtained by \citet{Krumholz2012} for the case of a constant gravitational field (i.e., vanishing gas self-gravity denoted by $g_\mathrm{gas} = 0$ in the legend; this limit can be obtained by setting
$f_{\rm gas} = 0$ in \autoref{eq_tempGrad} while maintaining finite optical depth).

Some trends evident in this figure are worth remarking on:
In the optically thin limit, the temperature gradient washes out and the atmosphere is increasingly well approximated as isothermal with a dimensionless temperature of $\Theta \to 1$.
This means that imposing balance of gravity and radiation pressure at infinity increasingly well corresponds to imposing balance of these forces all the way down to the midplane.
Note, however,  that the limiting $\fEc$ is gas fraction dependent: from \autoref{BC_3}, as the midplane temperature gradient washes out, 
the density profile must be increasingly well described as that for an isothermal atmosphere with a non-positive density gradient.
But then consideration of \autoref{eqn_MPdensGrad} shows that we must have $(1- f_\mathrm{\rm gas} - f_{E,*}) \geq 0$  in the $\tau \to 0$ limit.
This sets the critical, dimensionless Eddington ratio as $\fEc \to 1 - f_\mathrm{\rm gas}$ in the optically thin limit.
In the opposite, optically thick limit, the temperature gradient grows and the midplane temperature becomes $\gg 1$.
In this case, the radiation pressure gradient can render the atmosphere unstable towards the midplane even if the Eddington condition is satisfied at infinity.
Thus
$\fEc$ becomes increasingly small for increasing  optical depth but again with a limiting behaviour that is gas fraction dependent (though in the opposite sense to previous): 
because gas self-gravity vanishes in the midplane, as $f_\mathrm{\rm gas}$ is dialled upwards in the optically thick limit, the midplane density declines and, relative to small $f_\mathrm{\rm gas}$, the density gradient flattens off.
This reduces the optical depth near the midplane, lowering the temperature gradient and the resulting midplane radiation pressure gradient, rendering the atmosphere comparatively more stable with increasing  gas fraction case at fixed large $\tau_*$ and $\fE$.

\subsection{Convective stability}

An important point of difference between our setup and the constant gravitational field case investigated by \citet{Krumholz2012,Krumholz2013} 
is that we now incorporate gas self-gravity, which contributes a vanishing force in the midplane.
Indeed, with $ds/d\xi \propto \rho_\mathrm{\rm gas}$ and $\Theta \propto T$ (so that both $ds/d\xi$ and $\Theta$ are positive) and with our radiation source located in the midplane
(so that  $- d\Theta/d\xi$ is also positive at $z=0$), an
equilibrium solution may, on the basis of  \autoref{eqn_MPdensGrad}, turn out to have positive, zero, or negative density gradient ($d^2s/d\xi^2$) in the midplane.
On physical grounds, a configuration with an inverted density distribution seems likely to be unstable, a point we now examine in detail.

\subsubsection{Nature of instability and true stability criterion}

Though physical intuition might indeed suggest that an inverted density distribution be unstable, to be rigorous we should consider two questions: 
i) what is the nature of the (putative) instability? and ii) for what sort of configuration will the column be susceptible  to such instability?

With respect to question i), for the particular circumstances we investigate here, a number of potential instabilities can be immediately ruled out.
\citet{Blaes2003} find that local, radiative instability occurs only in the presence of magnetic fields or when the opacity contains an explicit density dependence, neither of which condition is met here. (In real galaxies magnetic fields are of course present, but we shall see below that another instability is more important in any event.)
The RRT instability described by \citet{Jacquet2011} is an interface instability that cannot develop for the smoothly-varying density distribution allowed  in the equilibrium situation we investigate.

In fact, we find that the putative equilibrium density and temperature profiles that correspond to equilibrium for $\fE$ approaching $\fEcr$ are susceptible to the classical convective instability.
The criterion for convective stability -- that the specific entropy, $S_g$, increase outwards -- may be expressed as the requirement that \citep{Blaes2003} the 
Brunt-V{\" a}is{\" a}l{\" a} frequency, $N_g$, be real or,
\begin{equation}
N_g^2 \equiv -\frac{(\gamma - 1)\rho T}{\gamma p_{\rm tot}} \mathbf{g} \cdot \nabla S_g > 0 \, ,
\end{equation}
where 
\begin{equation}
S_g \equiv \frac{k_B}{\mu (\gamma - 1)} \log\left(\frac{p_{\rm tot}}{\rho^\gamma} \right) + {\rm const} 
\end{equation}
is the specific entropy of the gas and $\gamma=5/3$ is the adiabatic index.\footnote{Note that it is solely the gas adiabatic index that appears in these equations.  While there are modes for which the radiation entropy profile matters, it is only the entropy of the gas that matters for the classical convective instability \citep{Blaes2003}. This holds even if the radiation contributes non-negligibly to the total energy budget. Also note that our adoption of a fixed $\gamma=5/3$ is something of an oversimplification for H$_2$. For a realistic ortho- to para-H$_2$ ratio of 3, $\gamma \approx 5/3$ at temperatures up to $\sim 50$ K, and decreases smoothly to $\gamma \approx 1.4$ over the temperature range from $\sim 50 - 500$ K \citep{Decampli1978, Boley2007}. Given that the difference between $\gamma = 5/3$ and $\gamma = 1.4$ only amounts to a $\approx 50\%$ change in the adiabatic temperature gradient we derive below, this effect is unlikely to be significant.
} 
From these equations, a sufficient condition for convective stability is $\mathbf{g} \cdot \nabla S_g < 0$ or (for our 1-D situation with $\nabla X = dX/dz \equiv X'$)
\begin{equation}
\frac{T'}{T + T_{\rm NT}} - \frac{(\gamma - 1)\rho'}{\rho} > 0 \, ,
\end{equation}
where $T_{\rm NT} \equiv T_* \Theta_{\rm NT}$ is a constant, dimensional, non-thermal temperature that accounts for turbulent energy density and pressure.
Rearranging and putting in terms of our dimensionless parameters, for convective stability we require
\begin{equation}
\frac{d \Theta  }{d \xi} \geq \left(\frac{d \Theta }{d \xi}\right)_{\rm ad}  \, , 
\label{eq_tempGradCS0}
\end{equation}
where
\begin{equation}
\left(\frac{d \Theta }{d \xi}\right)_{\rm ad} \equiv (\gamma - 1) \left(\Theta + \Theta_{\rm NT}\right) \frac{d^2 s}{d \xi^2}  \frac{d \xi}{d s}  \, .
\label{eq_tempGradCS}
\end{equation}
The subscript `ad' indicates that this is the value of $d\Theta/d\xi$ for which the gas is adiabatic. Combining this equation with \autoref{eq_mtmBalI} for hydrostatic equilibrium we find that the temperature gradient of a column that is simultaneously convectively stable and in equilibrium must be less negative than
\begin{equation}
\left(\frac{d \Theta }{d \xi}\right)_{\rm ad} = \frac{\gamma-1}{\gamma}\left[\fE k_r + (1- s) f_{\rm gas} -1 \right] \, .
\label{eq_tempGradCSII}
\end{equation}

For a column in equilibrium with radiation sources located in the midplane, $d\Theta/d\xi$ is minimised (most negative) in the midplane while $\Theta$ and $(1-s)$ are maximised in the midplane meaning that convective stability is most difficult to satisfy there; thus an equilibrium configuration that is convectively stable in the midplane is stable over the entire column.
On the other hand, for $\xi \to \infty$, $\left(d\Theta/d\xi\right)_{\rm ad} \to (\gamma-1)/\gamma \ (\fE - 1) $ which is less than zero for any putative hydrostatic equilibrium configuration (which must have $\fE < 1)$ while $d\Theta/d\xi \to 0$ (from below) so a hydrostatic system is always convectively stable at infinity. 
Most generally, at finite $\xi$ one can see that both increasing the fraction of total surface density in gas or dialling up the Eddington ratio at infinity, $\fE$, tend to render the system less stable with respect to convection.

\subsubsection{Convectively stable equilibria}

Under what circumstances will convective stability hold? To answer this question, we next identify the family of  curves (distinguished by their gas fraction $f_{\rm gas}$) giving the maximum value of the 
Eddington ratio at infinity $\fE$ (which we call $f_{\rm E,stab}$) that, at a given reference optical depth $\tau_*$, is simultaneously in mechanical equilibrium {\it and}  convectively stable.
To determine these curves it is sufficient to use the numerical procedure described above and then impose a refined constraint on the midplane density gradient 
that can be obtained from a rearrangement of \autoref{eq_tempGradCS}, namely, that:
\begin{equation}
\frac{d^2s}{d\xi^2} \leq \frac{1}{\gamma-1}\frac{d\Theta}{d\xi}\frac{ds}{d\xi}\frac{1}{\Theta + \Theta_{\rm NT} } \, .
\end{equation}
at $z=0$. 
We display curves for $f_{\rm E,stab}$ for the case $\Theta_{\rm NT} = 0$ with three representative values of $f_{\rm gas}$ (together with $\fEcr$ curves for the same $f_{\rm gas}$ values) in \autoref{fig_plotStabilityRegion}.

We can obtain an explicit form for $\fEs$
from a rearrangement of \autoref{eq_tempGradCSII} applied to the midplane, viz.:
\begin{equation}
 \fEs   = \left[\frac{1}{k_r} \left( 1-f_\mathrm{\rm gas} + \frac{\gamma}{\gamma-1}\frac{d\Theta}{d\xi} \right)\right]_{z=0}.
\end{equation}
This relation reveals that in the limit $f_\mathrm{\rm gas} \to 1$ a convectively stable column actually requires $d\Theta/d\xi > 0$ at $z=0$.
Such a positive midplane temperature gradient seems likely unphysical if the radiation sources are located there but, in the star formation context, the limit $f_\mathrm{\rm gas} \to 1$
is unphysical anyway given 
there has to be mass in the stars responsible for the  radiation field.

\subsection{Modified stability curves with convective heat transfer}
\label{sec:stab curves with cnvtn}

Thus far we have only considered hydrostatic equilibria where radiative transfer determines (via \autoref{eq_tempGrad}) the temperature gradient in a self-gravitating gaseous disc.
In \autoref{fig_plotStabilityRegion}, for a given value of $f_{\rm gas}$, the region of the parameter space between the $\fEs$ and the $f_{\rm E,flat,rt}$ curves represents a potential equilibrium configuration
which is, however, convectively unstable.
If convection is initiated over some range of the gas column, one must consider the possibility that convective heat transport will modify the temperature profile, exactly as it does in stars. However, the effects of convection in this case are subtle, because the situation is somewhat different than is typically found in stellar interiors. To see this, note that the heat flux per unit area carried by material convection must be of order
\begin{equation}
F_{\rm conv} \sim \rho \frac{k_B}{\mu} v_c \ell_c \left|\frac{dT}{dz}\right|,
\end{equation}
where $v_c$ is the characteristic convective velocity and $\ell_c$ is the characteristic size of a convective eddy. Non-dimensionalising this expression and normalising to $F_*$,
\begin{equation}
\frac{F_{\mathrm{conv}}}{F_*} \sim \frac{\tau_*}{\fE} \frac{c_{g,*}}{c} \mathcal{M} \ell  \left|\frac{d\Theta}{d\xi}\right|,
\end{equation}
where $c_{g,*} = \sqrt{k_B T_*/\mu}$ is the isothermal sound speed at temperature $T_*$, $\mathcal{M} = v_c/c_{g,*}$ is the Mach number of the convective flow, and $\ell = \ell_c/z_*$ is the convective Eddy size normalised to the isothermal scale height. The convective eddies cannot be much larger than the scale height, and $\Theta$ must drop from its maximum to unity over about a scale height, so the factor $\ell |d\Theta/d\xi|$ is at most of order unity. Similarly, for conventional convection the Mach number $\mathcal{M}$ cannot be large compared to unity. In contrast, observed galaxies possess photospheric temperatures, $T_*$, of, at most, $\sim 100$ K, in which case $c_{g,*}/c \sim 10^{-6}$. Thus even for relatively large values of $\tau_*$ and small $\fE$, we are still likely to have $F_{\mathrm{conv}}/F_* \ll 1$, indicating that conventional convection cannot contribute significantly to carrying the heat flux.

Convection in this radiation-dominated regime is a topic of current research. The most thorough numerical exploration to date is that of \citet{Jiang2015, Jiang2017}, who investigate convection in the diffuse outer layers of massive stars where, in analogy with our situation, the density is low enough that even transsonic matter motion does not carry a significant heat flux. They find that, if the optical depth per pressure scale height is smaller than the ratio of the gas sound speed to $c$ (as is the case for us), the gas becomes porous, allowing significantly larger radiative fluxes to pass through the matter than would be the case for a laminar matter distribution. In effect, the gas is convective, but the heat flux is carried by bubbles of radiation rather than hot matter. Particularly in the presence of magnetic fields this leads to a much flatter entropy profile than would be predicted for standard radiative transport, but not as perfectly flat as would be produced by efficient convection in a stellar core, for example.\footnote{The instability can also drive supersonic motions in the gas, which in principle should contribute to pressure support and thus set a minimum value of $\Theta_{\rm NT}$ in \autoref{eq_hydroBal}. However, we can ignore this complication because, while the turbulent pressure becomes larger than the gas pressure in this regime, it is always much smaller than the radiation pressure, and is therefore subdominant when it comes to determining hydrostatic balance.}

Given the uncertainty in the nature of this regime of radiation-dominated convection, and the paucity of numerical sampling of parameter space that would be used to calibrate a model based on mixing length theory or the like, our approach is simply to bracket reality by considering the two extreme limits. One limit is to assume that, in this regime, convection carries negligible heat flux. In this limit, we ignore the effects of convective instability and calculate the heat transport exactly as we would in the absence of convection (i.e., using \autoref{eq_tempGrad}) and $\fEcr$ sets the largest $\fE$ that permits an equilibrium. 
The opposite limit is to assume that convection is so efficient that it is, by itself, able to flatten the entropy gradient completely, as is usually the case deep in a stellar interior. Mathematically, this limit is equivalent to replacing $d\Theta/d\xi$ with $(d\Theta/d\xi)_{\rm ad}$ (\autoref{eq_tempGradCSII}) wherever solution of \autoref{eq_tempGrad} results in a value of $d\Theta/d\xi$ more negative than $(d\Theta/d\xi)_{\rm ad}$.

Operationally, we determine the profiles in the limit of efficient convection as follows. As we have seen above, an equilibrium column may become convective from the midplane up to some finite height $\xi_{\rm conv}$, above which radiative transfer again determines the temperature profile. Thus the equation governing the temperature profile over the height range $0 \to  \xi_{\rm conv}$ for a column that has attained marginal stability with respect to convection is given by \autoref{eq_tempGradCS0}, with the inequality replaced by an equality. For $\xi > \xi_{\rm conv}$ the temperature profile again becomes determined by radiation transfer (\autoref{eq_tempGrad}). The value of $\xi_{\rm conv}$ is determined implicitly by
\begin{equation}
\left(\frac{d\Theta}{d\xi}\right)_{\mathrm{rt},\, \xi=\xi_{\rm conv}} \equiv 
\left(\frac{d\Theta}{d\xi}\right)_{\mathrm{ad},\, \xi = \xi_{\rm conv}}
\label{eq_convSwitch}
\end{equation}
where the LHS of this equation is the temperature gradient due to laminar radiative transfer, and is defined by the (analytic) integral of \autoref{eq_tempGrad} subject to B.C.~\ref{BC_3}:
\begin{equation}
\left(\frac{d\Theta}{d\xi}\right)_{\mathrm{rad,}\, \xi=\xi_{\rm conv}} = -\frac{\tau_* k_r }{4 \lambda  \Theta^3 } \frac{d s }{d \xi} \, .
\end{equation}
At $\xi_{\rm conv}$ we require continuity of $\Theta$ and $s$ and their first derivatives, though the second and higher derivatives will be discontinuous in general. We can therefore solve the problem numerically by integrating \autoref{eq_hydroBal} and \autoref{eq_tempGradCS0} together starting from $\xi = 0$ until we reach a height where \autoref{eq_convSwitch} is satisfied; this defines $\xi_{\rm conv}$. We then switch to integrating \autoref{eq_hydroBal} and \autoref{eq_tempGrad} together from $\xi_{\rm conv}$ to infinity, using the values of $\Theta$, $s$, and their derivatives at $\xi_{\rm conv}$ as boundary conditions for this stage. We show an example density and temperature profile generated via this procedure in the solid lines in \autoref{fig_plotProfilesForVADERPaper1}.

\begin{figure}
\centering
\includegraphics[width = 0.5 \textwidth]{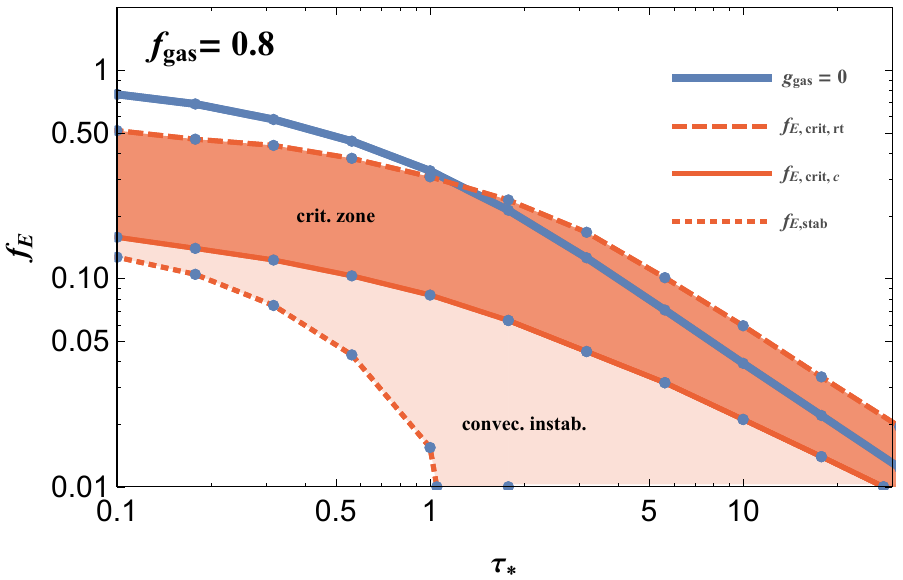}
\caption{Maximum $\fE$ such that a gas column with $f_{\rm gas} = 0.8$, $k_R = \Theta^2$, and $\Theta_{\rm NT} = 0$ can be in hydrostatic balance in the limiting cases that radiation-dominated convection is able to flatten the entropy gradient completely (solid line, $\fEcc$) and that radiation-dominated convection transports no more heat than laminar radiative transfer (dashed line, $\fEcr$). The true stability condition must lie in the heavily shaded band between the solid and dashed curves; in the lightly shaded zone the atmosphere is convectively unstable but otherwise hydrostatic. As in the previous figures, the solid blue line, for reference, is the vanishing gas self-gravity (fixed gravitational field) case investigated previously by \citet{Krumholz2012}.
} 
\label{fig_plotRTEqmRegion}
\end{figure}

\begin{figure}
\centering
\includegraphics[width = \columnwidth]{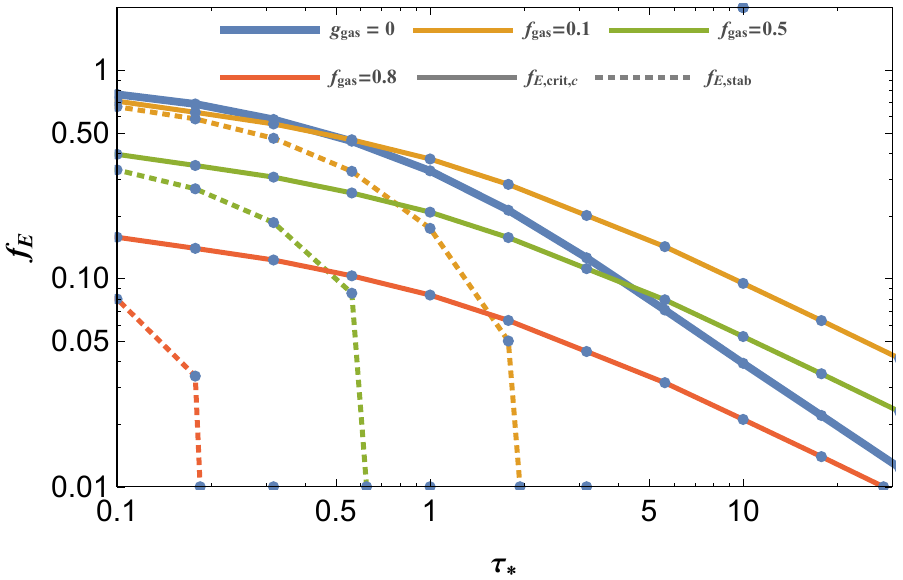}
\caption{
Curves for  $\fEcc$ (upper in each pair, solid) and $\fEs$ (lower  in each pair, dotted; see caption to \autoref{fig_plotStabilityRegion}) at different gas fractions with $k_R =  \Theta^2$ and $\Theta_{\rm NT} = 0$.
The upper curves correspond to the maximum $\fE$ such that a a gas column can be in hydrostatic balance (for the nominated gas fractions)
in the limiting case that radiation-dominated convection is able to flatten the entropy gradient completely.
} 
\label{fig_plotConvectiveStabilityRegion}
\end{figure}

Given this procedure for obtaining profiles in the limit of perfectly efficient convection, we can now repeat our analysis above to determine, for any specified gas fraction $f_{\rm gas}$, the value of $\fEcc$: the largest possible $\fE$ such that the dusty gas column can be hydrostatic. We show example results of this calculation in \autoref{fig_plotRTEqmRegion} for a gas fraction $f_{\rm gas} = 0.8$.
Here, the lowest curve indicates $\fEs$, the value of $\fE$ at which the gas first becomes unstable to convection. 
The upper curve is $\fEcr$, the maximum value of $\fE$ for which a hydrostatic atmosphere can exist if there is no convection and heat transfer is solely due to radiative transfer through a laminar medium. 
The middle curve indicates $\fEcc$, the maximum value of $\fE$ for which a (quasi-)hydrostatic atmosphere exists in the presence of effective convection.
In reality, where convection is neither perfectly efficient nor negligible, the true stability limit must lie between the two upper curves. The difference between the two curves is relatively modest even up to $\sim 50\%$ gas fractions, but can become large for even higher gas fractions. We illustrate how $\fEcc$, the stability limit assuming efficient convection, depends on gas fraction in  \autoref{fig_plotConvectiveStabilityRegion}.

\subsection{Stability curves with turbulence: the effect of non-zero $\Theta_{\rm NT}$}
\label{sec:ThetaNT}

In real, star-forming gas discs (the focus of \autoref{sec:implications}), gas is always highly turbulent as a result of supernova feedback and gravitational instability \citep[e.g.,][]{Krumholz2017}. We must therefore consider how such turbulence, driven by mechanisms other than radiation pressure, might modify the conditions under which radiation pressure can drive additional turbulence or eject gas in a wind. The interaction of turbulence with radiation when the gas is optically thin to infrared radiation (the so-called single scattering limit) has previously be investigated by \citet{Thompson2016}, but here we are interested in the case where the gas is optically thick to the infrared.

\begin{figure}
\centering
\includegraphics[width = \columnwidth]{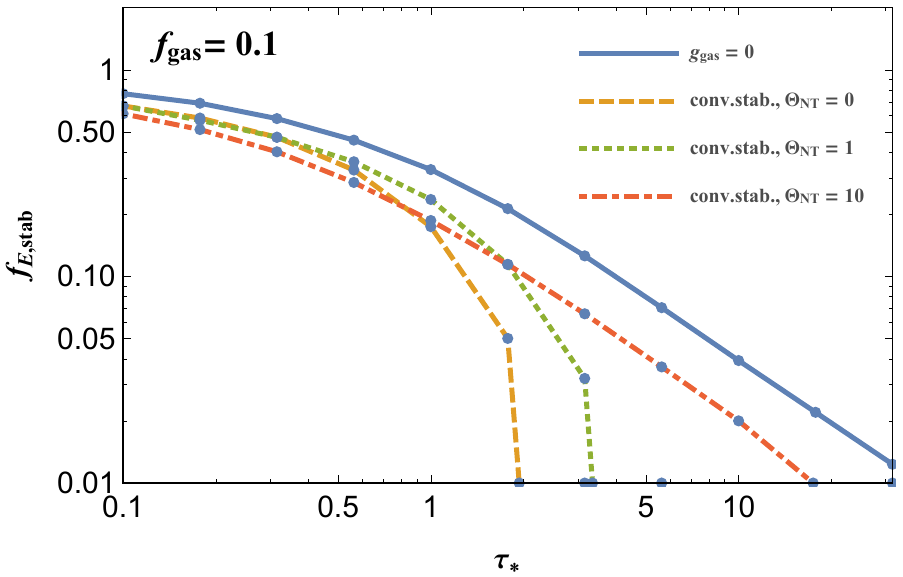}
\caption{Values of $\fE$ at which a gas column with $f_{\rm gas} = 0.1$ becomes convectively unstable for $\Theta_{\rm NT} = 0$, 1, and 10, as illustrated in the legend. As usual, we include the stability line for $g=\mbox{const}$ from \citet{Krumholz2012} for reference.
} 
\label{fig_plotConvectiveStabilityRegionII}
\end{figure}

While full solution of this problem will ultimately require simulations, we can qualitatively estimate the effects of turbulence by adopting a non-zero value for the non-thermal ``temperature" $\Theta_{\rm NT}$, which parameterises the degree of turbulent support. Qualitatively, the effect of non-zero $\Theta_{\rm NT}$ is to flatten the gas density profile and increase the scale height. Since the convective stability condition is critically-dependent on the sharpness of the temperature and density profiles, the primary effect of increasing $\Theta_{\rm NT}$ is to render the gas column more stable against convection. We illustrate this effect in \autoref{fig_plotConvectiveStabilityRegionII}, where we plot convective stability lines for different values of $\Theta_{\rm NT}$. Clearly the effect of non-zero $\Theta_{\rm NT}$ is to render the gas more convectively stable when the optical depth is high.

\begin{figure}
\centering
\includegraphics[width = \columnwidth]{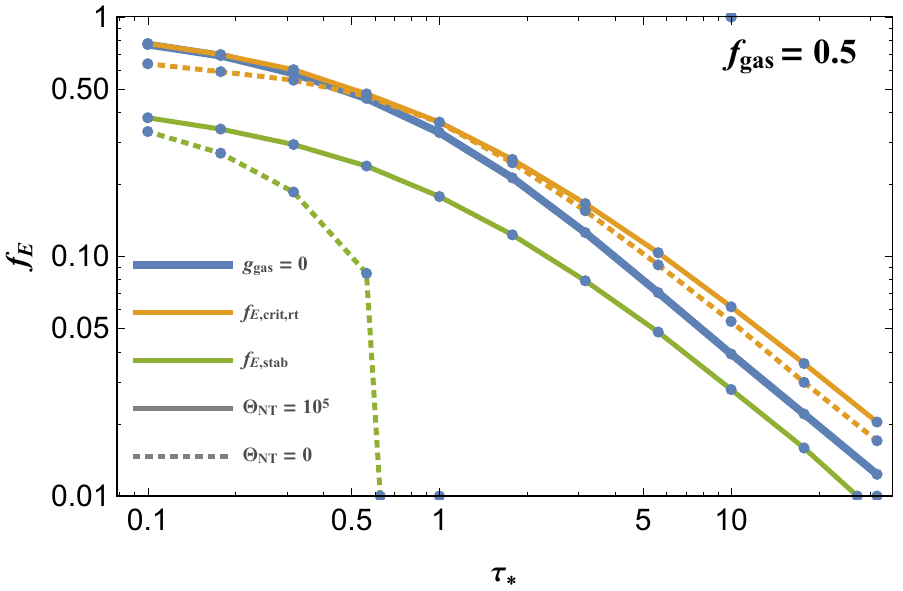}
\caption{Curves of hydrostatic stability assuming no convective heat transport $\fEcr$ (yellow) and convective stability $\fEs$ (green) for columns with no turbulence ($\Theta_{\rm NT} = 0$, dashed) and strong turbulence ($\Theta_{\rm NT} = 10^5$, solid). The blue line is the $g=\mbox{const}$ stability line of \citet{Krumholz2012} for reference.
} 
\label{fig_plotThetaNTCombined}
\end{figure}

While the inclusion of turbulent support alters the conditions under which the gas is convectively unstable, it has almost no effect on the critical value of $\fE$ at which it is no longer possible for the gas to be hydrostatic. We illustrate this in \autoref{fig_plotThetaNTCombined}, where we compare  stability lines with $\Theta_{\rm NT} = 0$ and $\Theta_{\rm NT} = 10^5$ for the example of a column with $f_{\rm gas} = 0.5$. (Our choice of $\Theta_{\rm NT}=10^5$ is explained in \autoref{ssec:dimensionless_to_physical}.) Clearly even for very large $\Theta_{\rm NT}$ the locus at which hydrostatic equilibrium becomes impossible if the assume that convection is ineffective ($\fEcr$) is essentially unchanged. Physically, we can understand this effect as resulting from the fact that, when radiation is the primary carrier of heat, the temperature profile depends only on the column density, and not the absolute height. Thus a large value of $\Theta_{\rm NT}$ increases the scale height of the atmosphere, but has no effect on the run of temperature versus column, or $\Theta$ versus $s$ in our non-dimensional variables. Since stability depends mostly on this relationship, the point at which stability is lost is mostly insensitive to $\Theta_{\rm NT}$. Moreover, because the $\fEcc$ curve must lie between $\fEs$ and $\fEcr$ (cf. \autoref{sec:stab curves with cnvtn}), and the effect of non-zero $\Theta_{\rm NT}$ is is simply to push $\fEs$ toward $\fEcr$, one can see that the real locus of hydrostatic equilibrium can only be very mildly dependent $\Theta_{\rm NT}$ and, thus, turbulence.

Put another way, turbulence does not affect stability if we assume that heat transport is dominated by radiation, because in this regime all turbulence does is make the atmosphere more extended without altering the relationship between temperature and gas column. If we consider the possibility that convection might transport heat, then the effect of turbulence is simply to push the convective case closer to the non-convective one, since anything that increases the gas scale height will flatten temperature gradients and thus reduce the tendency for convection to start up. Because heat transport is bounded between these two limiting cases, turbulence has little effect on stability.

\section{Implications for star-forming systems}
\label{sec:implications}

We have now determined, for a galactic disc of a specified gas fraction and photospheric optical depth $\tau_*$, the critical Eddington ratio $\fE$ below which the radiation is unable to set the gas into motion at all, and above which it is likely to eject it in bulk. Our next step is to translate this critical line in dimensionless space into the space of gas and star formation rate surface density, the observable quantities used most often to characterise star-forming systems.

\subsection{From dimensionless to physical quantities}
\label{ssec:dimensionless_to_physical}

To map our curves into this new parameter space, we begin by noting that, while observers often report star formation rates, the directly observable quantity is in fact the radiation flux in some tracer; for the starburst systems with which we are concerned, this tracer is generally the total infrared luminosity, which is taken as a proxy for the bolometric luminosity since most of the flux emerges in the infrared. This quantity is then converted to a star formation rate per unit area via a conversion factor:
\begin{equation}
F_* = \Phi \dot{\Sigma}_*.
\end{equation}
\citet{Kennicutt2012} recommend a conversion factor of $\Phi = 4.1\times 10^{17}$ erg g$^{-1}\approx 6.7\times 10^{9}$ $L_\odot$ $(M_\odot\mbox{ yr}^{-1})^{-1}$. Given this mapping between flux and observed areal star formation rate, we can immediately express the mapping between the dimensionless parameters $\tau_*$ and $\fE$ and the physical ones $\Sigma_{\rm gas}$ and $\dot{\Sigma}_*$. 
Using the opacity scaling $\kappa_R = \kappa_0 (T/T_0)^2$, and the definitions of $T_*$, $\fE$, and $\tau_*$ given above we can define
\begin{equation}
F_{\rm *,crit} \equiv \frac{\left(16\pi G c \sigma_{\rm SB} T_0^4\right)^{1/2}}{\kappa_0 } \simeq 1.9 \times 10^{13} \lsun \ {\rm kpc}^{-2} \, ,
\end{equation}
which is nearly identical to the limit derived by \citet{Thompson2005} in their consideration of self-gravitating optically-thick disks.
Then, after some algebra one can show that
\begin{eqnarray}
\dot{\Sigma}_* & = & \frac{F_{\rm *,crit}}{\Phi}  \sqrt{\frac{\tau_* \fE}{f_{\rm gas}}} 
\nonumber \\
& = & 2.8\times 10^3 \sqrt{\frac{\tau_* \fE}{f_{\rm gas}}} \,M_\odot\mbox{ yr}^{-1}\mbox{ kpc}^{-2}
\label{eq_SigmaSFR}
\end{eqnarray}
and
\begin{eqnarray}
\Sigma_{\rm gas} & = & \frac{2}{\sqrt{\kappa_0}} \left(\frac{\sigma_{\rm SB} T_0^4}{\pi G c}\right)^{1/4}
\left(\frac{\tau_*^3}{f_{\rm gas}^3 \fE}\right)^{1/4} 
\nonumber \\
& = & 5.3\times 10^3 \left(\frac{\tau_*^3}{f_{\rm gas}^3 \fE}\right)^{1/4}\,M_\odot\mbox{ pc}^{-2}.
\label{eq_SigmaGas}
\end{eqnarray}
For the numerical evaluations we have used $\kappa_0 = 10^{-1.5}$ cm$^2$ g$^{-1}$ and $T_0 = 10$ K\footnote{Note that in the presence
of an additional source of radiative flux connected to an AGN, $F_\mathrm{AGN}$,
the RHS of \autoref{eq_SigmaSFR} for the critical  star formation rate surface density contains an additional term $- F_\mathrm{AGN}/\Phi$ whereas \autoref{eq_SigmaGas} is unmodified.}, appropriate for the dust abundance in the Solar neighbourhood. We specialise to this case because, as we show below, trapped radiation pressure will prove to be important only in the most luminous and actively star-forming galaxies, and these are invariably observed to be near-Solar in their metallicities. However, it is trivial to extrapolate the results to non-Solar metallicities, since examination of the above equations immediately reveals that, at fixed $\tau_*$ and $f_{\rm E,*}$, changes in the value of $\kappa_0$ simply scale the star formation and gas surface densities as $\dot{\Sigma}_* \propto \kappa_0^{-1}$ and $\Sigma_{\rm gas} \propto \kappa_0^{-1/2}$. Thus the effect of varying the dust opacity per unit mass is simply the slide the stability curves that we derive below along a line of slope 2 in the $(\Sigma_{\rm gas}, \dot{\Sigma}_*)$ plane.

We pause here to remark on two related issues.
First, note that below we use critical curves calculated in \autoref{sec:equilibria} with vanishing non-thermal temperature (i.e., $\Theta_{\rm NT} = 0$)
for determining the stability of real, star-forming systems.
This seems to neglect the point that
gas in real, star-forming systems is riven by extrinsic, non-thermal velocity dispersion imposed by both 
gravitational instability and supernova feedback and this turbulence hugely puffs up the gas in such systems.
Moreover, the scale of $\Theta_{\rm NT}$ is large: let us parameterise the turbulent pressure as
\begin{equation}
P_{\rm turb} \simeq \rho  \sigma_{\rm gas}^2 \equiv  \frac{ \rho}{\mu} k_B  T_* \Theta_{\rm NT} \, .
\end{equation}
Then, note that we expect that real systems self-regulate so as to achieve a
 $Q_{\rm gas}$  close to 1 \citep[][and references therein]{Krumholz2017,Thompson2005} where
\begin{equation}
Q_{\rm gas} = \frac{\kappa \sigma_{\rm gas} }{\pi G \Sigma_{\rm gas}} \, ,
\end{equation}
in which $\kappa$ is the epicyclic frequency.
Taking $\kappa \simeq \Omega \sim 200$ km/s/kpc (where $\Omega$ is the angular frequency), typical for a ULIRG, we normalise the gas velocity dispersion in such a system to
\begin{equation}
\sigma_{\rm gas} \simeq 340 \ {\rm km/s} \ Q_{\rm gas} \left(\frac{\Sigma_{\rm gas}}{5000 \ \msun/{\rm pc}^2}\right) \left(\frac{\kappa}{200 \ {\rm km/s/kpc}} \right)
\end{equation}
so that in the most extreme systems we expect
\begin{equation}
\Theta_{\rm NT} = 1.4 \times 10^5 \left(\frac{ \sigma_{\rm gas}}{340 \ {\rm km/s}}\right)^2 \left(\frac{T_*}{100 \ {\rm K}}\right)^{-1} \, .
\end{equation}
%
However, as far as the stability of real systems subject to indirect radiation pressure goes, this discussion is moot: as we have already shown in \autoref{sec:ThetaNT},  the introduction of a large $\Theta_{\rm NT} \sim 10^5$ does not materially change  the true, critical $\fE$ above which hydrostatic equilibrium cannot be attained (which lies between $\fEcc$ and $\fEcr$).
Nor will it change the midplane temperature or consequent radiation energy density.
In other words, the critical locus for $\fE$, and the consequent radiation pressure stability curves in the Kennicutt-Schmidt parameter space we have calculated above, 
all carry through to realistic cases where high degrees of turbulence puff up star-forming, gaseous discs so that they are Toomre stable. 

A second important point is that, 
up to dimensionless constants of $O(1)$, $F_{\rm *,crit}$ is identical to the characteristic flux identified by \citet{Thompson2005} \citep[also cf.][]{Scoville2003}
for marginally Toomre-stable ($Q \sim 1$), optically thick, star-forming discs radiating at their Eddington limit.
However, nowhere above have we assumed $Q \sim 1$; 
in fact, we have shown that the conditions for hydrostatic equilibrium to be possible are nearly independent of $\Theta_{\rm NT}$, and thus of $Q$. 
In fact, the direct correspondence between $F_{\rm *,crit}$ and the characteristic flux previously derived by  \citet{Thompson2005} is a result of the fact that the limit does not depend on the vertical gas density distribution or its scale height but only on the overall optical depth.

Returning to our main argument,  \autoref{eq_SigmaSFR} and \autoref{eq_SigmaGas} allow us to translate a curve in the $(\tau_*, \fE)$ plane directly into one in the $(\Sigma_{\rm gas}, \dot{\Sigma}_*)$ plane, provided we know the gas fraction. While this is sometimes also directly observable, in many instances it is not, particularly for the starburst systems of greatest interest to us; in these galaxies, high dust columns can make it almost impossible to observe the old stellar population, particularly at high redshift. For this reason, it is helpful to consider what gas fractions are possible at a given point in $(\Sigma_{\rm gas}, \dot{\Sigma}_*)$-space. At any given point in this space the gas fraction is bounded from above, because there is minimum mass in stars required to produce the observed light. The light to mass ratio of a simple stellar population with a standard IMF has a maximum value $\Psi \approx 2200$ erg s$^{-1}$ g$^{-1} \approx 1100$ $L_\odot$ $M_\odot^{-1}$ \citep{Fall2010}, and declines thereafter. Thus an ``observed" star formation rate $\dot{\Sigma}_*$ (in reality an observed bolometric flux $F_* = \Phi \dot{\Sigma}_*$) requires a minimum stellar mass per unit area $\Sigma_* = (\Phi/\Psi)\dot{\Sigma}_*$ to produce it. The gas fraction therefore has a maximum value
\begin{equation}
f_{\rm gas,max} = \frac{\Sigma_{\rm gas}}{\Sigma_{\rm gas} + (\Phi/\Psi) \dot{\Sigma}_*}.
\end{equation}

More generally, it is convenient to express the mapping between position in the $(\Sigma_{\rm gas}, \dot{\Sigma}_*)$ plane using the approximation suggested by \citet{Krumholz2010}. They point out that, for a stellar population formed by continuous star formation over a time $t$, the light to mass ratio can be written approximately as
\begin{equation}
\frac{F_*}{\Sigma_*} \approx \frac{\Psi}{\max\left(1, t/t_{\rm cr}\right)},
\label{eq_lightToMass}
\end{equation}
where $t_{\rm cr} = \Phi/\Psi \approx 6.9$ Myr. The physical basis for this approximation is that for $t \ll t_{\rm cr}$ none of the massive stars producing the bulk of the light have had time to evolve off the main sequence and die, so the bolometric luminosity is simply proportional to the mass of the stellar population. For $t \gg t_{\rm cr}$ the massive stellar population reaches statistical equilibrium between new stars forming and older ones dying, and thus the luminosity becomes proportional to the star formation rate; since the stellar mass is just the star formation rate multiplied by $t$, the light to mass ratio therefore scales as $1/t$ for large $t$. Our \autoref{eq_lightToMass} simply interpolates between these two limits, with the value of $t_{\rm cr}$ chosen to ensure that $F_* \to \Phi \dot{\Sigma}_*$ as $t\to\infty$. Using \autoref{eq_lightToMass}, we can express the gas fraction at a given point in the $(\Sigma_{\rm gas}, \dot{\Sigma}_*)$-plane as a function of the effective stellar population age as
\begin{equation}
f_{\rm gas} = \frac{\Sigma_{\rm gas}}{\Sigma_{\rm gas} + \max(t, t_{\rm cr}) \dot{\Sigma}_*}.
\end{equation}
Equivalently, we can say that a specified gas fraction corresponds to a particular line of slope unity in the $(\Sigma_{\rm gas}, \dot{\Sigma}_*)$-plane,
\begin{equation}
\dot{\Sigma}_* = \left(\frac{1-f_{\rm gas}}{f_{\rm gas}}\right) \frac{\Sigma_{\rm gas}}{\max(t,t_{\rm cr})}.
\label{eq_lightToMassII}
\end{equation}

\subsection{Stability region for star-forming systems}

\begin{figure}
\includegraphics[width=\columnwidth,angle=90]{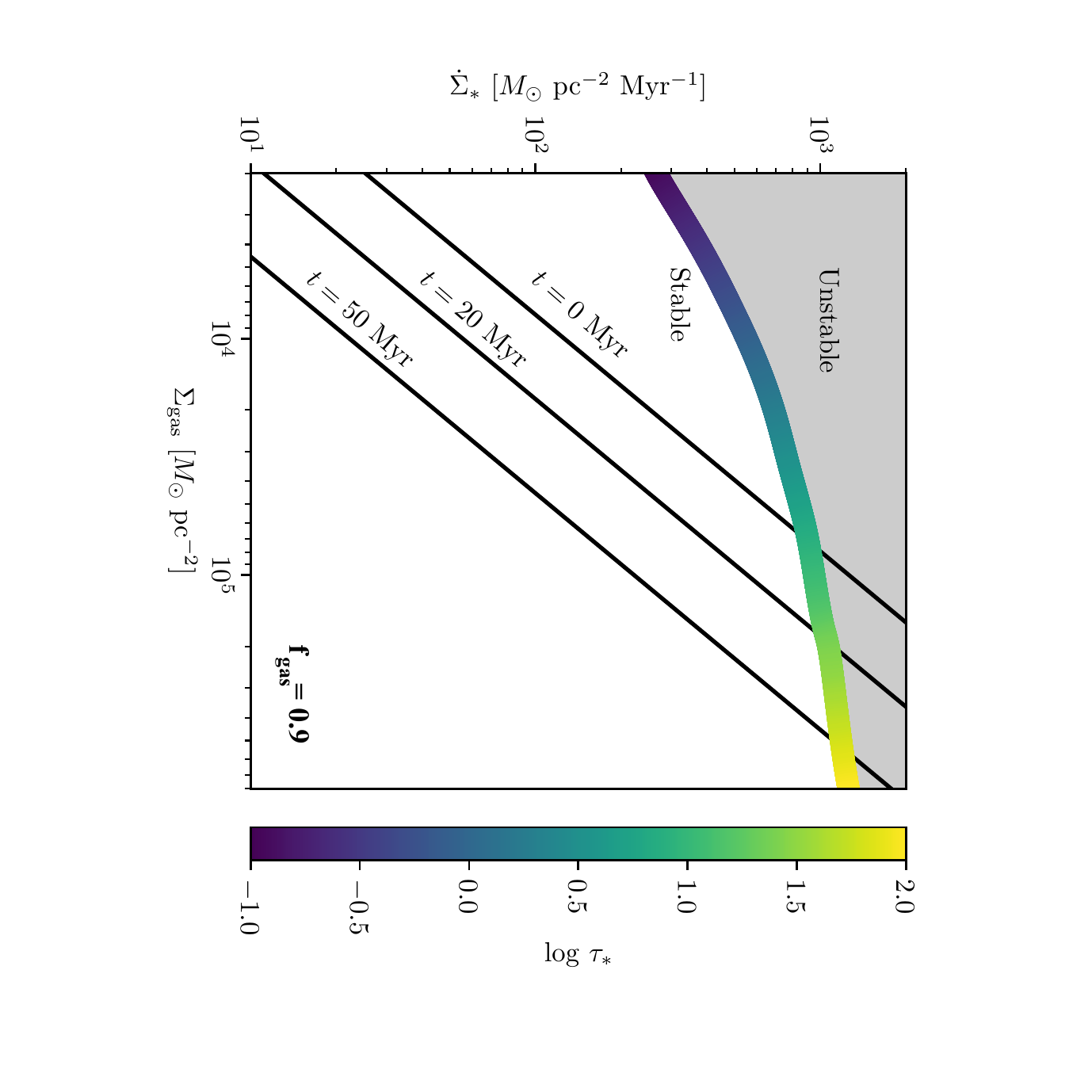}
\caption{
\label{fig:solution_diagram}
Schematic of the method for constructing the stability curve. Thin black lines show lines of constant gas fraction (\autoref{eq_lightToMassII}; in this example $f_{\rm gas} = 0.9$) for stellar population ages $t=0$, 20, and 50 Myr, as indicated. The thick coloured line indicates the stability curve (in this example we have used $\fEcc(\tau_*, f_{\rm gas})$, the stability curve assuming efficient convection) translated into $(\Sigma_{\rm gas}, \dot{\Sigma}_*)$ coordinates via \autoref{eq_SigmaSFR} and \autoref{eq_SigmaGas}; colour along the line indicates the value of $\tau_*$ at that point. For any choice of stellar population age, the gas and star formation rate surface density at which a hydrostatic solution ceases to exist for this gas fraction corresponds to the point where the thin black and thick coloured lines cross. In the unshaded region of parameter space labelled ``Stable", radiation pressure is unable to destablise the gas and cause a loss of hydrostatic balance, while in the shaded, ``Unstable" region it can.
Note that, in the presence of an additional source of radiative flux connected to an AGN, $F_\mathrm{AGN}$,
the stability curve would be shifted downwards by an amount $\sim F_\mathrm{AGN}/\Phi$.
}
\end{figure}

We are now ready to determine the locus of the stability curve in the  plane of observables. Mathematically this locus is defined by the solution to the non-linear system defined by \autoref{eq_SigmaSFR}, \autoref{eq_SigmaGas}, \autoref{eq_lightToMassII}, and the dimensionless stability curve $\fE = \fEc(\tau_*, f_{\rm gas})$. For any specified $t$ and choice of $f_{\rm gas}$, this constitutes a set of four equations in the four unknowns $\tau_*$, $\fE$, $\dot{\Sigma}_*$, and $\Sigma_{\rm gas}$, which is straightforward to solve numerically. Conceptually, one can visualise the solution procedure as shown in \autoref{fig:solution_diagram}. Choose a stellar population age $t$ and a value of $f_{\rm gas}$. Via \autoref{eq_lightToMassII}, this defines a line of slope unity in the $(\Sigma_{\rm gas}, \dot{\Sigma}_*)$-plane along which the solution must lie. Similarly, for fixed $f_{\rm gas}$, if one varies $\tau_*$ then \autoref{eq_SigmaSFR} and \autoref{eq_SigmaGas} define a parametric curve in the $(\Sigma_{\rm gas}, \dot{\Sigma}_*)$-plane, which represents the locus of stability. The point $(\Sigma_{\rm gas}, \dot{\Sigma}_*)$ where this curve crosses the constant gas fraction line is the combination of gas and star formation rate surface density that is marginally stable for the chosen gas fraction and stellar population of age.

By varying $f_{\rm gas}$, one traces out a curve in the $(\Sigma_{\rm gas}, \dot{\Sigma}_*)$-plane that defines the boundary between stable and unstable for all possible gas fractions at the chosen stellar population age; values of $\Sigma_{\rm gas}$ or $\dot{\Sigma}_*$ below this line are stable, those above it are unstable. This procedure can be applied for any stability curve of the form $\fE(\tau_*, f_{\rm gas})$, and thus we can use it to generate the curves where convection sets in, and where hydrostatic balance is lost under the limiting assumptions of maximally inefficient and maximally efficient radiation-dominated convection.

\begin{figure*}
\includegraphics[width = 0.75 \textwidth]{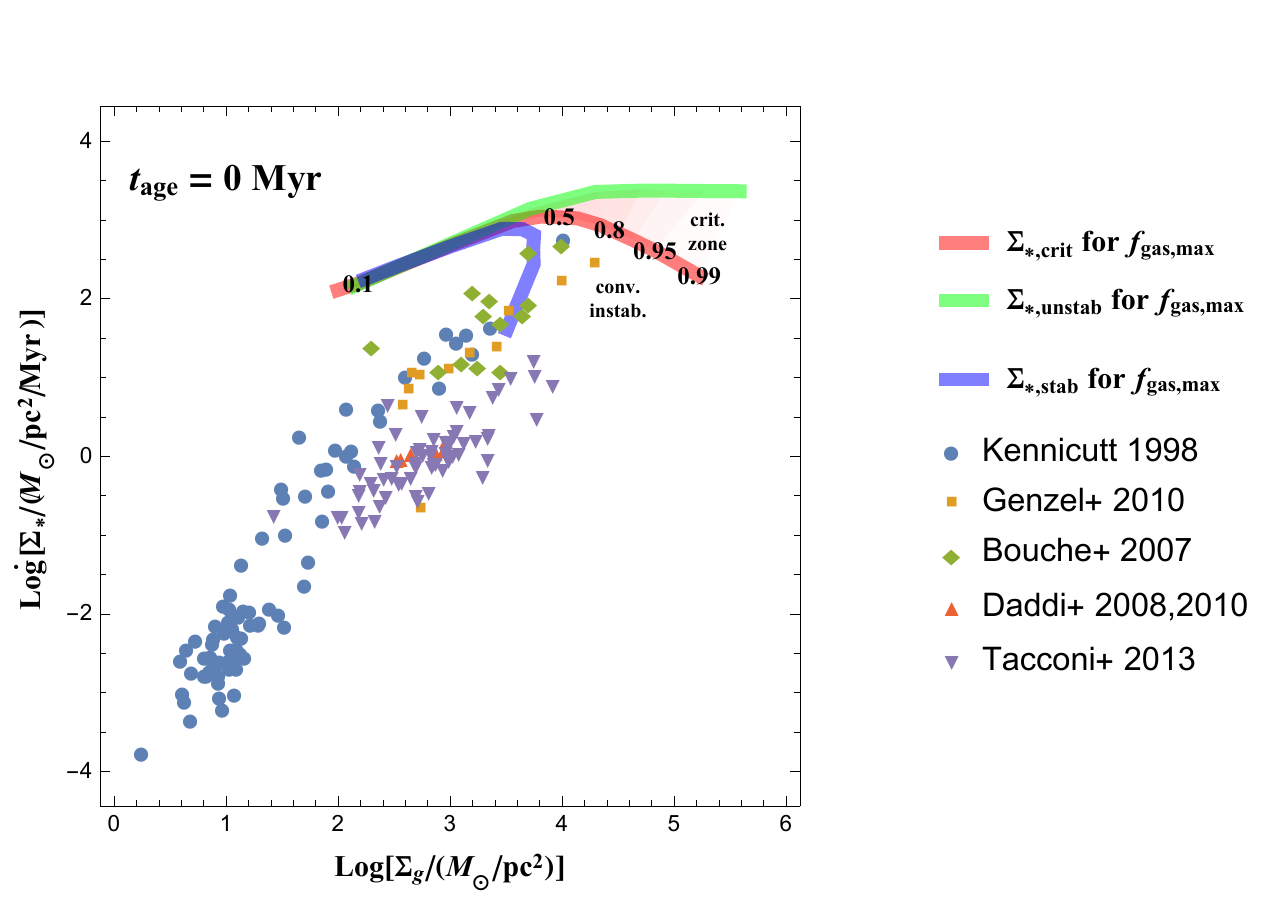}
\includegraphics[width =  0.75 \textwidth]{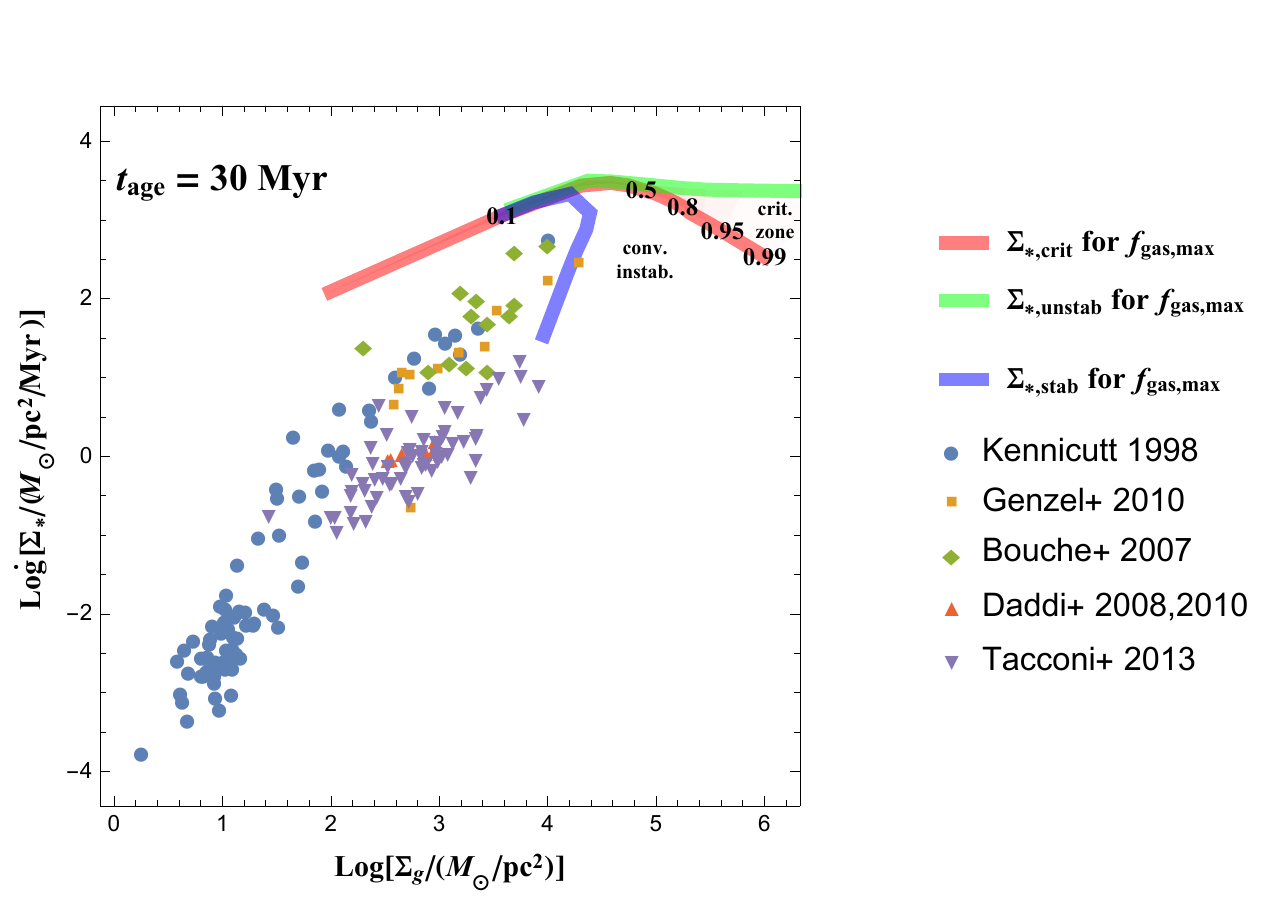}
\caption{
Radiation pressure stability curves versus observations. The top panel is for a stellar population age $t < t_{\rm cr} \approx 6.9$ Myr, while the bottom is for a stellar population age of 30 Myr. 
In both panels, the thick red curve shows the maximum star formation rate (for given gas surface density) for which a gas column can be in hydrostatic balance, assuming that radiation-dominated convection is able to flatten the entropy gradient completely; the value of the gas fraction along this curve is indicated in boldface numbers at selected points. The thick green curve shows the same quantity computed in the limit that radiation-dominated convection transports no more heat than laminar radiative transfer. The true stability condition must lie between these limits, in the region labeled critical zone. The thick blue curve shows the locus of convective stability; systems above and to the right of this curve are subject to convection, while those below and to the left are convectively stable. Coloured points show observed global star formation rates versus gas surface densities in a sample culled from the following sources: local galaxies from \citet{Kennicutt1998}, z $\sim 2$ sub-mm galaxies from \citet{Bouche2007}, and galaxies on and somewhat above the star-forming main sequence at z $\sim 1-3$ from \citet{Daddi2008,Daddi2010b,Genzel2010,Tacconi2013}.  The observations have been homogenised to a \citet{Chabrier2005} IMF and the convention for $\alpha_{\rm CO}$ suggested by \citet{Daddi2010a}; see \citet{Krumholz2012b} for details. 
}
\label{fig_plotSFRdataVsModel}
\end{figure*}

In \autoref{fig_plotSFRdataVsModel}, we show the loci of stability for two stellar population ages $t=0$ (though it would be identical for any $t < t_{\rm cr}$) and $t = 30$ Myr. In this figure, the red line shows the stability curve under the assumption that radiation-dominated convection is maximally efficient and flattens the entropy gradient perfectly, while the green line shows the stability condition under the assumption that a convectively unstable region transports no more heat than a stable one where the flux is carried by radiation alone. The true stability curve must lie between these two limits, in the shaded region marked ``critical zone'' in the plot. The blue curve shows the locus where convective instability occurs. In all cases the stable region is below and to the left of the curves, while the unstable region is above and to the right. 

In the Figure we also show a selection of observed galaxies culled from the literature. The primary point to take from this comparison is that the stability curves are generally far from the data, even in the most optimistic case where the stellar population age is assumed to be $\lesssim 10$ Myr (though we emphasise that, in this case, the star formation rate conversion that is normally adopted is invalid, and the star formation rate shown becomes merely a lower limit). For a more realistic but still optimistic case of a stellar population age of 30 Myr, not a single observed galaxy lies in region where radiation pressure prevents the atmosphere from being hydrostatic, even if we adopt the most optimistic assumptions about convection. Crucially, however, the stability curve is a surprisingly good match for the \textit{upper envelope} of the observed distribution, an observation whose implications we tease out in the next section.

\section{Discussion and Conclusion}
\label{sec:discussion}

The idea that radiation pressure on dust-bearing gas may be responsible for launching galaxy scale outflows dates back more than fifty years \citep{Harwit1962,Chiao1972,Ferrara1990}.
Furthermore, the role and importance of infrared radiation pressure as an agent of feedback in star cluster and galaxy formation has been
a subject of particularly intense scrutiny over the last decade
\citep{Scoville2003,Murray2005,Thompson2005,Fall2010, Murray2010,Murray2011,Krumholz2012,Krumholz2013,
Thompson2016, Raskutti2016}, and numerical simulations that treat radiative transfer with varying levels of sophistication have yielded sharply divergent results, with some finding that radiation pressure feedback is important in rapidly star forming systems \citep[e.g.,][]{Hopkins2011,Hopkins2012}, while others have found the opposite result \citep[e.g.,][]{Rosdahl2015b}. In this context, consideration of \autoref{fig_plotSFRdataVsModel} 
reveals one significant positive finding and one significant negative finding with respect to the possible role of indirect radiation pressure in regulating galaxy formation. We first focus on the latter, leaving the former to the following section.

\subsection{Infrared radiation pressure as a regulator of star formation}

Our negative finding is that the large majority of real star forming systems lie well within the region where radiation pressure is dynamically unimportant, and this remains true even if we focus solely on starburst galaxies that are far from the star-forming main sequence. Moreover,  the shape of the critical curve imposed by radiation forces in the $(\Sigma_{\rm gas}, \dot{\Sigma}_*)$-plane is not morphologically similar to the Kennicutt-Schmidt relation. This result, while consistent with some more recent studies (e.g.., \citealt{Reissl2018}), stands in contrast to at least some earlier work.
For instance, \citet{Thompson2005} determined that as systems transition from being optically thin to optically thick to reradiated infrared, their self-regulated, marginally-stable ($Q \sim 1$)
star-formation activity undergoes a corresponding transition in scaling from $\dot{\Sigma}_* \propto \Sigma_{\rm gas}^2$ to $\dot{\Sigma}_* \sim$ const; 
this scaling \citep[and the absolute normalisation of the relations determined by][]{Thompson2005} generated a plausible match to the empirical Kennicutt-Schmidt relation \citep{Andrews2011}.
Other authors have 
computed a dust Eddington limit by adopting a constant ``characteristic'' infrared opacity that is then held fixed \citep[e.g.,][]{Hopkins2010}, or using a simple scaling based on an estimated midplane temperature. A fixed IR opacity corresponds to a critical curve that is a line of slope unity in the $(\Sigma_{\rm gas}, \dot{\Sigma}_*)$-plane, again close enough to the slope of the observed Kennicutt-Schmidt relation to suggest a possible correspondence.

Here we have improved on these approaches by properly solving the equation of radiation transfer and thus determining the self-consistent run of density and temperature versus height implied by simultaneous radiative and hydrostatic balance. This in turn allows us to compute the true, self-consistent value of $\tau_{\rm IR}$. Our more accurate calculation shows that the true Eddington limit line bears little resemblance to the observed Kennicutt-Schmidt relation, and thus cannot be responsible for setting it, or for regulating star formation more broadly. However, this does not preclude that indirect radiation pressure effects may be important for regulating intense, localised star-formation on sub-galactic scales, i.e., in individual giant molecular clouds collapsing to form star clusters; we will revisit this question in future work. Nor does it rule out the possibility that radiation pressure effects might reduce the star formation rate by pressurising the ISM \citep[e.g.,][]{Rosdahl2015b, Costa2017}, though this seems unlikely to occur except quite near the gas ejection line, since this line is defined by the condition that radiation pressure begin to dominate the midplane.

From the observational side, the molecular gas in those few galaxies that may fall within the convectively unstable zone will, as already mentioned, be highly turbulent as a result of supernova feedback and gravitational instability \citep[e.g.,][]{Krumholz2017}. There is no obvious route to separating convective motions from turbulent ones, nor is there any reason to believe that any radiation-driven convective motions will be significant compared to those induced by gravitational instabilities or supernova explosions.

A corollary of this finding applies to numerical simulations and the subgrid models they employ, which also often rely on the \textit{ansatz} of a fixed infrared opacity. Recall that a central finding of radiation-hydrodynamic simulations to date is that for Eddington ratios below the critical value, radiation does not cause any gas motions or drive any turbulence.\footnote{Formally we note that we have found the regime where hydrostatic atmospheres exist, not proven that those atmospheres are stable. We can rule out the possibility of local instabilities in this regime, since none of the local instability conditions found by \citet{Blaes2003} are satisfied. However, we cannot completely rule out the possibility that  our hydrostatic atmospheres are subject to a heretofore undiscovered global instability, though the fact that no evidence of such an instability has emerged from the numerical simulations strongly suggests that this is not the case.} The reason this happens is that, in the stable regime, the gas column is able to self-adjust so that it settles to an opacity profile $\kappa(z)$ whereby at every point the outward radiative and pressure forces balance the inward gravitational force. A key part of this self-adjustment occurs through the temperature-dependence of the opacity, which provides a feedback loop between the density distribution and the radiative force: as the density distribution changes, the temperature profile and the radiative force do as well.

Now consider what happens when we remove this feedback loop by fixing $\kappa_{\rm IR}$ as in the subgrid models; for simplicity in this thought experiment, we will hold $g$ constant as well, though including self-gravity would lead to qualitatively the same conclusion. With fixed $\kappa_{\rm IR}$ and $g$, the ratio of gravitational to radiative force is constant, and we can immediately see that a wind will be driven whenever the flux $F_* > g c / \kappa_{\rm IR}$, or, in terms of our dimensionless variables, $\fE > \kappa_{R,*}/\kappa_{\rm IR}$. For $\kappa_{\rm IR} = 5$ cm$^{2}$ g$^{-1}$ (as used, for example, in \citealt{Hopkins2011}) and our standard opacity function (\autoref{eq:opacity_scaling}) and scaling between flux and star formation rate (\autoref{ssec:dimensionless_to_physical}), this condition evaluates numerically to $\fE > 7\times 10^{-3} \dot{\Sigma}_{*,0}^{1/2}$, where $\dot{\Sigma}_{*,0}$ is the star formation rate per unit area measured in units of $M_\odot$ pc$^{-2}$ Myr$^{-1}$. Thus for areal star formation rates typical of those found in high$-z$ galaxies, the constant $\kappa_{\rm IR}$ model predicts the launching of winds at Eddington ratios as small as $\sim 0.01$, independent of $\tau_*$. Comparison of this prediction to the true stability curves derived in \autoref{sec:equilibria} shows that Eddington ratios this small should lead to wind launching only for $\tau_* \gtrsim 10$, whereas most real galaxies have $\tau_* \lesssim 1$ (see \autoref{sec:tiring}). Thus a constant $\kappa_{\rm IR}$ model, at least for commonly-used values of $\kappa_{\rm IR}$, makes launching radiation-driven outflows much easier than it should be. The ultimate source of this problem is the choice to adopt a fixed opacity, rather than one that self-adjusts as a function of Eddington ratio and optical depth as it should. Whether incorrect wind launching actually occurs in any given simulation will depend on the distribution of Eddington ratios within it, which will in turn depend on the details of the local gravitational field and stellar luminosity. However, the fact that the use of a fixed $\kappa_{\rm IR}$ can easily lead to gas ejection in situations where it should not be possible is a source of concern for the results derived with current subgrid models. To avoid this problem, absent a simulation having the resolution and physics sufficient to capture the run of temperature versus position within an irradiated gas column, a second-best solution would be to explicitly estimate $\fE$ and $\tau_*$, and to inject enough momentum to drive a wind only if the condition $\fE>\fEc$ is met.

\subsection{Infrared radiation pressure as a limit to gas densities and star formation rates}
\label{sec:PIRlimit}

While our results imply that radiation pressure is not an important regulator of star formation in most galaxies on global scales, we have also found that the extremum in the $(\Sigma_{\rm gas}, \dot{\Sigma}_* )$ parameter space occupied by real systems is coincident with the critical line above which trapped radiation pressure turns on and is able to eject gas. We emphasise that there is no a priori reason why our calculation should have produced this result. In the dimensionless parameter space of $\tau_*$ and $\fE$ that defines our system, the critical value of $\fE$ above which gas is ejected follows purely from the mathematical form of the equations; the only astrophysical input to this result is the scaling of opacity with temperature, $\kappa \propto T^2$, which holds simply because the mean grain size is much smaller than the radiation wavelength. The translation of this line into the observational parameter space of gas and star formation surface density depends only on fundamental constants, on the light-to-mass ratios of stellar populations, and on the specific opacity of interstellar dust at Solar metallicity. Thus in our calculation of a critical Eddington ratio line, we have used no information whatsoever about galaxies or their assembly history.
The fact that our calculated limit nevertheless closely matches the observed upper limit on surface densities of star formation seems unlikely to be a coincidence, and strongly hints that ejection by indirect radiation pressure prescribes the region of the $(\Sigma_{\rm gas}, \dot{\Sigma}_*)$-plane that may be occupied. 
Indeed, it was this correspondence that motivated previous work on the importance of radiation pressure in extreme systems \citep{Thompson2005}.

The observed galaxies that come closest to the radiation pressure limit are recent merger systems like Arp 220 or sub-mm galaxies. Our finding suggests a scenario whereby mergers can drive gas to higher gas surface densities and star formation rates along the Kennicutt-Schmidt relation (which is set by physics that have little to do with radiation pressure), but if the surface density or star formation rate becomes too high, the system crosses the critical line. At that point radiation is suddenly able to eject the majority of the mass over a very short timescale, driving the surface density and star formation rate back down and to the left on the Kennicutt-Schmidt plot. Thus infrared radiation pressure sets a maximum flux for star-forming galaxies.

\section*{Acknowledgements}

The authors gratefully acknowledge conversations with Geoff Bicknell, Yuval Birnboim, and Chris McKee. MRK acknowledges support from the Australian Research Council's Discovery Projects grant DP160100695. TAT is supported in part by NSF \#1516967 and NASA 17-ATP17-0177.








\begin{appendices}

\section{Photon tiring}
\label{sec:tiring}

Radiation from a massive star may induce a quasi-steady-state wind whose mechanical luminosity is ultimately bound by the radiative luminosity at the wind base; this limit on the mechanical power of the wind is known as the photon tiring limit \citep[][]{Owocki1997,Owocki2004}. In our stability calculation we implicitly assume that, if the radiation flux is so large that no hydrostatic configuration is possible, the radiation will eject gas. However, this is possible only if the act of driving such a wind would not exceed the photon tiring limit, i.e., if it is possible to drive off a wind without using more energy than the radiation field has available. This situation is somewhat different than the case of a massive star where the wind is quasi-steady, but the analogous question of a galactic disc is whether there is sufficient power in the radiation field for it to remove the atmosphere within a dynamical time.

In the spirit of deriving the most stringent possible limit on when photon tiring will become important, 
we neglect gas internal energy and turbulence and consider only gas kinetic and gravitational and potential energy surface density. For a slab of gas that is being ejected at velocity $\dot{z}$, the energy content is therefeore
\begin{equation}
E_\mathrm{gas} = T_\mathrm{gas} + U_\mathrm{gas} \simeq \Sigma_\mathrm{gas} \left( \ \dot{z}^2/2 +  g z \right)
\end{equation}
To accelerate the gas upward in the potential well with an acceleration $\ddot{z}$, the rate per unit area at which the radiation field must do mechanical work on the gas is
\begin{equation}
\dot{E}_\mathrm{gas} \simeq \Sigma_\mathrm{gas} \dot{z} \left( \ddot{z} +  g \right).
\end{equation}
Again, in the spirit of deriving a lower limit, we consider the power required to raise the gas steadily ($\ddot{z} \to 0$) and take the ratio of this to the power per unit area in the radiation, viz. the radiative flux $F_*$.
The rough figure-of-merit, therefore, is
\begin{eqnarray}
\frac{\dot{E}_\mathrm{gas}}{F_*}  & \sim &   \frac{\Sigma_\mathrm{gas} \ \dot{z} \ g}{F_*}   \\
& =  &  \frac{\Sigma_\mathrm{gas} \ \dot{z} \ \kappa}{c \fE} 
 = \frac{\dot{z}}{c} \frac{\tau_*}{\fE} \simeq 10^{-3}  \frac{\tau_*}{\fE} \left( \frac{\dot{z}}{300 \ \mathrm{km \,s^{-1}}} \right) \, \nonumber .
 \end{eqnarray}
From this expression and by inspection of, e.g., \autoref{fig_plotRTEqmRegion}, 
the mechanical power per unit area required to eject the gas at a speed comparable to the escape speed from a galactic disc only approaches the radiative flux, $\dot{E}_\mathrm{gas}/{F_*} \sim 1$, for $\tau_* \gsim 30$. For comparison, using our Solar neighbourhood dust opacity (\autoref{eq:opacity_scaling}) and the \citet{Kennicutt2012} scaling between flux and star formation rate (\autoref{ssec:dimensionless_to_physical}), we have
\begin{equation}
\tau_* = 7.2\times 10^{-6} \left(\frac{\Sigma_{\rm gas,1/2}}{M_\odot\,\mathrm{pc}^{-2}}\right) 
\left(\frac{\dot{\Sigma}_*}{M_\odot\,\mathrm{pc}^{-2}\,\mathrm{Myr}^{-1}}\right)^{1/2}.
\end{equation}
Consulting \autoref{fig_plotSFRdataVsModel}, the highest observed gas surface densities and areal star formation rates are of order $10^4$ $M_\odot$ pc$^{-2}$ and $10^3$ $M_\odot$ pc$^{-2}$ Myr$^{-3}$, respectively, corresponding to $\tau_* \approx 2$. Thus observed galaxies are well away from the range where photon tiring is an important limit.

Moreover, our calculation of photon tiring neglects wind acceleration and, more importantly, relies on what is probably an unrealistically large normalising wind velocity.
While it may seem reasonable to normalise
 $\dot{z}$ to a circular velocity $v_\mathrm{circ}$ typical for a large spiral galaxy, the VET and IMC radiation hydrodynamics studies reviewed 
 in the \autoref{sec:intro} show that the gas atmosphere suffers a very mild, near-logarithmic acceleration in the super-Eddington case. When this happens the gas is ejected without the need for it to be accelerated to speeds comparable to the circular velocity.
In summary, it is safe to neglect photon tiring over the parameter space we consider.


\end{appendices}

\bsp	
\label{lastpage}
\end{document}